\documentclass[%
 aip,
 amsmath,amssymb,
preprint,%
nofootinbib,
]{revtex4-1}

\usepackage{tabularx}
\usepackage{graphicx}
\usepackage{dcolumn}
\usepackage{bm}

\usepackage[utf8]{inputenc}
\usepackage[T1]{fontenc}
\usepackage{mathptmx}
\usepackage{xcolor}
\usepackage{physics}
\usepackage{hyperref}
\usepackage{caption}
\usepackage{subcaption}

\UseRawInputEncoding
\def\pmb#1{\setbox0=\hbox{#1}
	\kern-.025em\copy0\kern-\wd0
	\kern.05em\copy0\kern-\wd0
	\kern-.0125em\raise.0433em\box0 }

\def\bx {\pmb {$x$}}
\def\bc {\pmb {$c$}}
\def\bu {\pmb {$u$}}
\def\bf {\pmb {$f$}}
\def\by {\pmb {$y$}}

\def\bR {\pmb {$R$}}

\begin{document}

\preprint{}

\title[]{Rotational and Reflectional Equivariant Convolutional Neural Network for data-limited applications: Multiphase Flow demonstration}

\author{B. Siddani}
 \affiliation{Center for Compressible Multiphase Turbulence, University of Florida, Gainesville, FL 32611, USA}
\author{S. Balachandar}%
 \email{bala1s@ufl.edu}
\affiliation{Center for Compressible Multiphase Turbulence, University of Florida, Gainesville, FL 32611, USA}%

\author{R. Fang}
\affiliation{J. Crayton Pruitt Family Department of Biomedical Engineering, University of Florida, Gainesville, FL 32611, USA}%

\date{\today}

\begin{abstract}
This article deals with approximating steady-state particle-resolved fluid flow around a fixed particle of interest under the influence of randomly distributed stationary particles in a dispersed multiphase setup using Convolutional Neural Network (CNN). The considered problem involves rotational symmetry about the mean velocity (streamwise) direction. Thus, this work enforces this symmetry using \textbf{SE(3)-equivariant}, special Euclidean group of dimension 3, CNN architecture, which is translation and three-dimensional rotation equivariant. This study mainly explores the generalization capabilities and benefits of SE(3)-equivariant network. Accurate synthetic flow fields for Reynolds number and particle volume fraction combinations spanning over a range of [86.22, 172.96] and [0.11, 0.45] respectively are produced with careful application of symmetry-aware data-driven approach.
\end{abstract}

\maketitle
\section{Introduction}

Availability of big data and high computing power through ever advancing technology has drawn immense attention in recent times toward the application of data-driven techniques in the field of fluid mechanics. Extremely successful applicability of these methods to various fluid mechanics problems is the main reason behind their evident popularity. \citet{Brunton_ML_FM} discussed many current and emerging developments of Machine Learning (ML) in fluid mechanics. A significant number of the current ML applications are related to turbulence modeling. Different aspects of data-driven turbulence modeling have been presented in a review by \citet{turb_ml}. The challenges associated with ML-augmented sub-grid models is discussed in [\onlinecite{duraisamy2021perspectives}]. Abstract representation of flow physics, when sufficient data is available, and ease of implementation have made Neural Networks (NNs) the first choice among many ML techniques for fluid applications. The abstract nature of neural networks also suggests that prior known information such as governing equations of the flow and symmetries present in the flow are not generally enforced.  

\citet{RAISSI2019686} introduced Physics-Informed Neural Networks (PINNs) which explicitly enforce the governing equations in the loss function. This ensures that solutions of neural networks satisfactorily obey the governing equations. This approach has shown improved generalization capabilities of the trained networks. Application of PINN approach to high-speed flows has been studied in [\onlinecite{MAO2020112789}]. A similar approach of including residues of governing equations in the loss function was implemented by \citet{subramaniam2020turbulence} to perform super-resolution of an incompressible forced isotropic homogeneous turbulence problem, and by \citet{jiang2020meshfreeflownet} to super-resolve turbulent flows in the Rayleigh-B\'ernad convection problem. Of relevance to the present work is the recent work of \citet{siddani2020machine} who used a generative adversarial network (GAN) to predict the flow around a particle in a particle-laden flow given information on the local mean flow and the relative location of its neighbors.

As seen in the above listed applications, the primary use of neural network in fluid mechanics has been to serve as models, which when trained well are designed to accurately predict the desired output quantities based on an input feature vector. For example, in a turbulence closure model, given the local strain-rate tensor of the resolved scale flow, the objective of the NN is to evaluate the subgrid stress due to the unresolved scales. In a super-resolution problem, given the coarse resolution flow on a plane or a volume as input, the NN is trained to predict a highly resolved flow within the same plane or volume. In the work of [\onlinecite{siddani2020machine}], the Convolutional Neural Network (CNN) was trained to predict the entire velocity and pressure fields within an array of randomly distributed particles only from information on their position and particle Reynolds number. 

It is well known that such models must obey certain underlying symmetry and conservation principles, some of which are listed below. i) {\it{Translational invariance:}} also known as spatial homogeneity, implies that if the input features were to be specified at two different points, planes, or volumes that are shifted by a spatial translation, then the predicted flow quantities in both these situations must be identical. ii) {\it{Rotational Equivariance:}} implies that if the input features were to be specified at two different points, planes, or volumes that differ by a rotation, then the predicted flow quantities in the two situations must be related through the same rotation. Note that under rotation, scalar quantities will remain identically the same (invariance), whereas vector and tensor quantities are only related and will not be identically the same (equivariance) iii) {\it{Reflectional equivariance:}} implies that if the input features were to be specified at two different points, planes, or volumes that differ by a reflection, then the predicted flow quantities must be related through the same reflection. iv) {\it{Galilean invariance:}} implies that if input features were to be specified in two different frames of reference that differ by a uniform velocity, then the predicted flow quantities in two frames of reference must be identical. v) {\it{Kinematic conditions:}} such as boundary conditions at interfaces or the incompressibility condition must be satisfied by the predicted flow quantities. vi) {\it{Conservation laws:}} of mass, momentum, and energy must also be satisfied by the predicted flow quantities to the extent that they are applicable.

Let us consider the above list of NN requirements in the context of predicting the flow around a reference particle surrounded by a random distribution of neighbors. In this context, translational invariance says that the predicted flow should not depend on where the reference particle is located within a particle-laden flow as long as its neighborhood remains the same. 3D (three-dimensional) CNN, such as the ones used in \citet{siddani2020machine}, by design, satisfy translational invariance along all three directions. Similarly, rotational and reflectional equivariances imply that if the neighborhood around the reference particle were to be arbitrarily rotated or reflected from an initial configuration, the corresponding predicted flows must be accordingly rotated or reflected versions of the solution predicted in the initial configuration. Galilean invariance can be ensured by requiring that the input and output features of the network include only relative velocity information between the reference particle and the ambient flow and between the different particles. The general approach to enforcing kinematic conditions and conservation equations has been to calculate them for the predicted solution and any error in their satisfaction will be applied as an additional loss term in the training of the NN. 

Thus, translational invariance, Galilean invariance, kinematic conditions, and conservation equations are satisfied in a CNN by design, as part of the loss function, or with a careful choice of input and output feature vectors. In comparison, enforcement of rotational and reflectional symmetries has generally been in an approximate sense through data augmentation, which tends to increase the cost of the training process. As a result, machine learned models have been sensitive to arbitrary coordinate system used in their training process [\onlinecite{smidt2020euclidean}]. This suggests the importance of building-in the desired rotational and reflectional symmetries into ML models by carefully designing their architecture. For example, \citet{wang2020incorporating} incorporated rotational symmetry into data-driven models for two-dimensional (2D) fluid flow systems and their results showed improved accuracy. Development of equivariant neural networks which strictly enforce different forms of symmetry is an active area of research.

\textit{The quest of this article is to present a 3D rotationally (SO(3), special orthogonal group of dimension 3) equivariant CNN and demonstrate its ability to predict fluid quantities at greater accuracy, especially under conditions of paucity of training data.} Furthermore, it will be shown that the resulting symmetry-constrained network is more compact and involves optimization of far fewer network weights during the training process. In fluid mechanical applications, the requirement of rotational equivariance is in the context of a mixture of scalar (e.g., pressure), vector (e.g., velocity), pseudo-vector (e.g., vorticity) and tensor (e.g., strain-rate, stress) quantities, that can be part of the input features or the predicted output. Enforcement of rotational eqivariance for a mixture of scalar, vector, and tensor quantities requires planning and will be discussed. These features of equivariant CNN will be discussed for the prediction of fluid flow around a particle of interest (reference particle) under the influence of its neighboring particles with a strict enforcement of rotational symmetry about the mean-flow direction. The importance and need for such symmetry preserving networks in the limit of scarce data will be reasoned and demonstrated through quantitative results of the considered multiphase flow problem. The discussed rotationally equivariant approach is an efficient alternative to data augmentation process of considering arbitrary rotations which is a necessity in applications with data paucity.     

The remainder of this paper is arranged as follows: Section II describes the problem in detail. Section III presents details of the dataset used in the demonstration of SE(3)-equivariant, special Euclidean group of dimension 3, CNN. Section IV provides an overview of SE(3)-CNN. The machine learning approach is detailed in section V. The performance and properties of a trained network are evaluated in section VI. Finally, the carried out study is summarized in section VII. This section also briefly mentions some future possibilities of the presented methodology.

\section{Problem Description}
\begin{figure}[h]
    \centering
    \includegraphics[width=\linewidth,keepaspectratio=true]{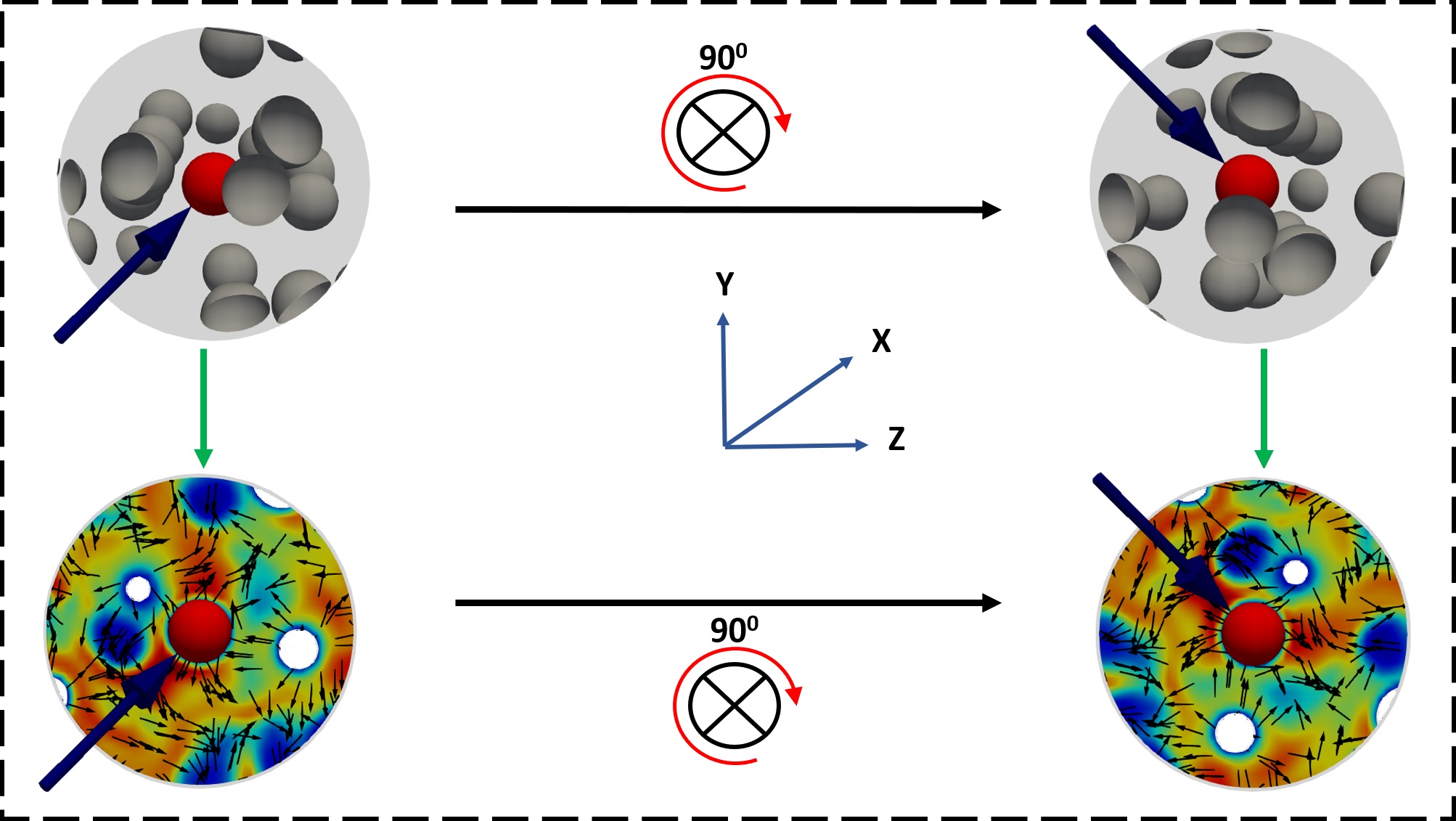}
    \caption{Visual depiction of equivariance of flow systems.}
    \label{fig:equivariance}
\end{figure}

Consider a large random distribution of particles in which one particle is arbitrarily chosen as the reference particle, shaded red as shown in Figure \ref{fig:equivariance}. The neighbors of the reference particle within a sphere of five diameters in size are shown shaded in gray in Figure \ref{fig:equivariance}. Given the average fluid velocity within this sphere in a frame attached to the reference particle (indicated by the velocity vector), here we are interested in predicting the flow around the reference particle, which is influenced by the presence of neighbors. As shown in \citet{siddani2020machine} the velocity and pressure fields within a considered volume are first predicted which can then improved in the region close to the reference particle.

An example of such a velocity field plotted on the central $y-z$ plane of this domain is shown in the bottom left corner. The contour represents $u$ (velocity component along $x$ direction) and the arrows represent the in-plane 2D vector created using $v$ \& $w$, velocity components along $y$ \& $z$ respectively. The white circular patches of different sizes correspond to the cross-sections of neighboring particles that intersect with this plane. For the sake of further discussion this domain will be called \textit{original configuration} in the current discussion. 

Now consider a new particle configuration referred to as the \textit{rotated configuration}, shown in the upper right hand side of the figure, which corresponds to the `original configuration' rotated about the $x$-axis by $90^\circ$ in the clockwise direction. Note that the average flow velocity within the sphere, which is also part of the input feature along with the location of the neighboring particles, is also correspondingly rotated. The velocity field of the `rotated configuration' in the central $y-z$ plane is shown in the bottom right image of the figure. Rotational equivariance implies that the resulting velocity field in the rotated configuration must be related to that of the original configuration by a rotational transformation, which is evident in Figure \ref{fig:equivariance}. As the axis of rotation in this example is $x$-axis, the velocity component $u$ contours are simply rotated about the center, while the in-plane velocity vectors undergo vectorial transformation.  

The present work makes use of 3D Steerable Convolutional Neural Network (CNN) developed by \citet{weiler20183d}. 3D Steerable CNNs, which operate on regular 3D grids, are equivariant to 3D rotations and translation - i.e., these neural networks are SE(3)-equivariant to 3D rotations and translations. Since the translational invariance property of a regular CNN is well established [\onlinecite{weiler20183d}] here we will discuss mainly the structure of the network that enforces rotational equivariance (i.e., SO(3)-equivariance). We note that in Figure \ref{fig:equivariance} rotation about the $x$ axis is presented only as an example, and the equivariance equally applies to rotations about any other axis. In the remainder of this paper such a CNN will be referred to as SE(3)-CNN. 

Generalization capabilities of a fairly complex neural network increase with the amount of training data. Data augmentation opportunities are often exploited to increase the training dataset size, since the amount of raw 3D data available for training from fluid mechanical simulations and experiments is not usually very large. Performing discrete rotations and reflections of the training data is a common method to increase the total sample size [\onlinecite{tempogan,siddani2020machine}]. The number of discrete rotations that can be performed as part of data augmentation can be arbitrarily high. However, the total training time of the network increases with increasing data augmentation. The inherent SE(3)-equivariance property of a SE(3)-CNN ensures that it achieves the desired generalization capabilities without actually performing data augmentation through discrete rotations. In fact, an SE(3)-CNN enforces continuous 3D rotations, and hence, enjoys a superior generalization capability than that obtained through discrete rotational data augmentation procedure.

The present focus on predicting the flow around the reference particle is inspired by the need to predict the force on the particle more accurately taking into account the fluid-mediated influence of its neighbors. There have been others ML efforts that have explored different networks to obtain the force on a particle. \citet{HE2019379} used artificial neural network (ANN) to predict drag force on a particle in a randomly distributed monodispersed, stationary, spherical particles system. The inputs to the ANN were Reynolds number ($Re$) of the flow, particle volume fraction ($\phi$) and relative positions of 15 nearest neighbours, and the drag force being the required output. \citet{PhyNet} improved this approach through the incorporation of pressure and velocities at discrete points around the particle's surface, shear and pressure forces acting on the particle as outputs of network's intermediate layers. These physically-meaningful intermediate variables were optimized through the inclusion of mean squared error (MSE) terms with respect to corresponding PR-DNS values in the loss function. \citet{MOORE2019187} formulated a method that combined the pairwise-interaction extended point-particle (PIEP [\onlinecite{akiki2017pairwise}]) framework with a data-driven approach, based on nonlinear regression, to obtain particle forces. 

The superposition of physics-based PIEP model and data-driven approach enabled this method to achieve accurate predictions for both low and high particle volume fraction cases. Furthermore, \citet{superposablewake} introduced the \textit{superposable wake} method which like the present SE(3)-CNN also attempts to predict the flow around a reference particle under the perturbing influence of neighbors. While the SE(3)-CNN simultaneously takes into account the influence of all the neighbors within the surrounding neighborhood, the superposable wake method accounts for the effect of neighbors taken one at a time using pairwise superposition. Nevertheless, the physics-based superposable approach by construction automatically satisfies the SE(3) symmetry of the problem. Whereas, in the case of CNN, SE(3)-equivariance is not automatically guaranteed and must be enforced by special construction of the network. An important takeaway from all of the above mentioned works is that it is vital or even mandatory to include flow physics into ML models with the inclusion of all symmetries, invariances and conservation laws in order to get high quality results.

\section{Training and testing data}
The present objective is to recreate the particle resolved flow around a particle of interest under the influence of its neighbors using SE(3)-CNN when the geometric location of all particles, volume-averaged Reynolds number ($Re$) and mean flow direction are given as input information. These stated input quantities are readily available in a standard Euler-Lagrange simulation [\onlinecite{balachandar_eaton}] The neighboring particles are randomly distributed across the entire domain with uniform probability and the reference particle is considered at the center of the domain. Furthermore, all particles are spherical in shape with same diameter (monodispersed) and they are also considered to be stationary. It is worth mentioning that particle volume fraction ($\phi$) for the domain can be calculated as the location and total number of particles are given. 

\begin{figure}[h]
    \centering
    \includegraphics[width=\linewidth,keepaspectratio=true]{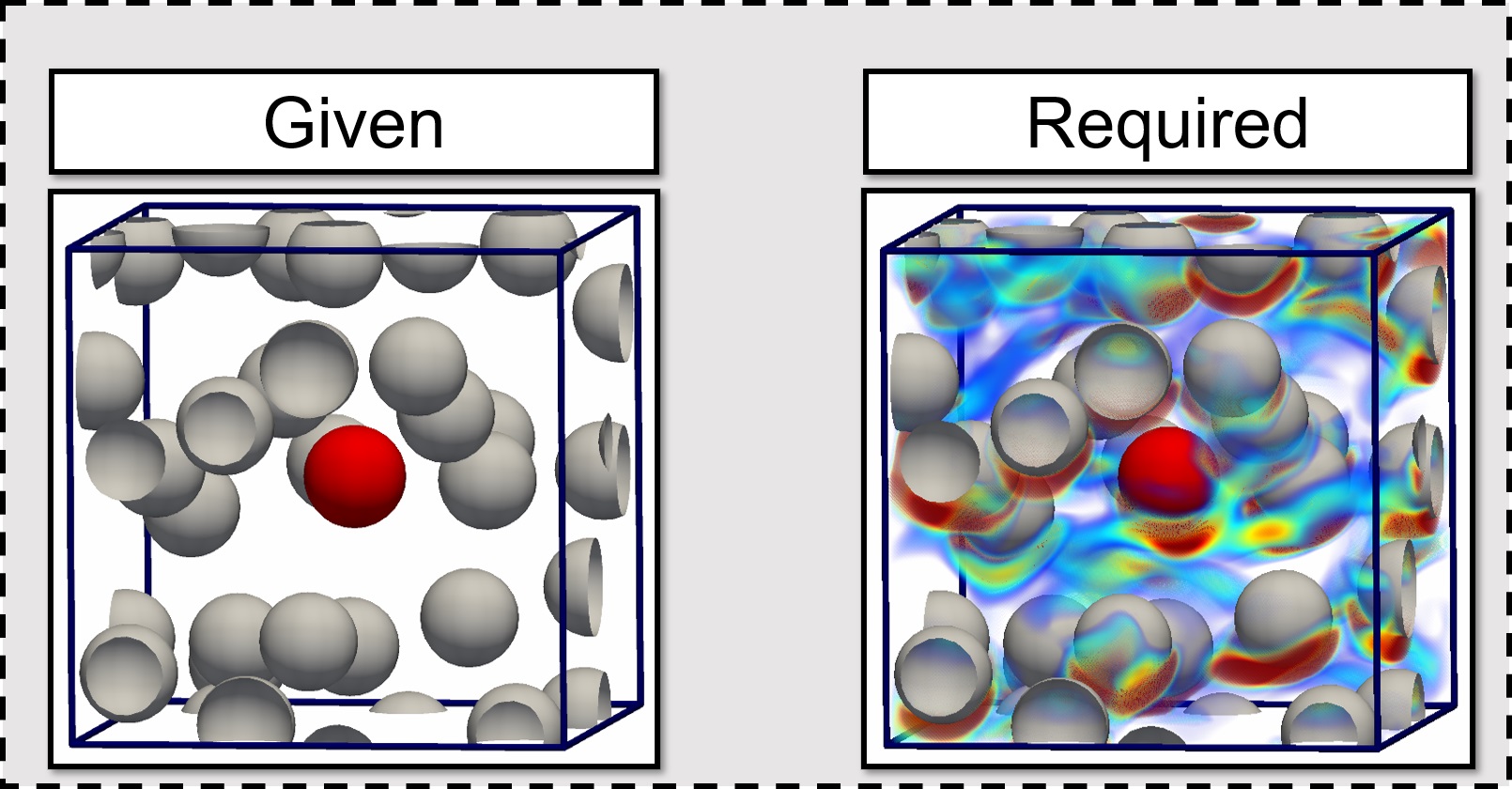}
    \caption{\textit{Objective}: Predict flow fields around a particle of interest under the influence of its neighbors when the volume-averaged Reynolds number, mean flow direction, and location of the particles are given.}
    \label{fig:objective}
\end{figure}

\subsection{Datasets}

Particle-resolved direct numerical simulation (PR-DNS) data is used to train and test the network. The current work utilizes the results presented in [\onlinecite{MOORE2019187,AKIKI_IBM}]. These simulations are performed in a cubic domain. The diameter of particles $d_{*}$ is chosen as the length scale for non-dimensionalization, thereby making the non-dimensional diameter equal to unity. The non-dimensional length of each side of the cubic domain is $3\pi$. The domain is subjected to periodic boundary conditions along $x$ and $y$ directions. Symmetric, no-stress boundary conditions are applied along $z$ direction. Pressure gradient is applied such that the mean flow (streamwise direction) of the entire computational domain is along the $x$ direction, although the mean flow averaged over smaller boxes around each particle could slightly deviate from this entire domain average. Non-dimensional form of the incompressible Navier-Stokes equations are solved to obtain the PR-DNS data. The different simulated cases can be classified using the two macroscale parameters, namely, particle volume fraction ($\phi$) and particle Reynolds number ($Re$). Table~\ref{tab:cases} shows all the cases considered in this study.

\begin{table*}
    \begin{ruledtabular}
			\begin{tabular}{ccccccccccc}
				Case&$Re$ & $\phi$ & Realizations & $\sigma_u$ & $\sigma_v$ & $\sigma_w$ & $\sigma_p$ & $N_{tr}$ & $N_{val}$ & $N_{te}$\\
				\hline
				1&172.96 & 0.11 & 10 & 0.4968 & 0.1947 & 0.1960 & 0.1973 & 372 & 46 & 46 \\
				2&86.22 & 0.21 & 7 & 0.5960 & 0.2580 & 0.2609 & 0.3807 & 450 & 90 & 90 \\
				3&114.60 & 0.45 & 5 & 0.8006 & 0.4340 & 0.4421 & 1.0467 & 579 & 193 & 193 \\
			\end{tabular}
		\end{ruledtabular}
		\caption{The Reynolds number and volume fraction of the different cases considered, the corresponding RMS values of $u, v, w, p$ and the size of training ($N_{tr}$), validation ($N_{val}$) and testing ($N_{te}$) datasets.}
		\label{tab:cases}
\end{table*}

\subsection{Data Preprocessing}
The analysis of perturbation force maps obtained using the superposable wakes (for Table~\ref{tab:cases} cases) by \citet{Balachandar-PIEP} indicate that the influence of neighbors that are three or four diameters away from the reference particle is practically zero. This supports the concept of considering a local box, which encompasses the region of highly influential neighbors, around the reference particle to obtain the flow field around it. Here, we consider a local box of non-dimensional size $5 \times 5 \times 5$ with the reference particle at its center and the flow within it is resolved with a grid of $64 \times 64 \times 64$ points. This consideration of local volume is the same as that used in \citet{siddani2020machine} and the process of creating these data samples from PR-DNS [\onlinecite{MOORE2019187,AKIKI_IBM}] has been thoroughly detailed in it. The total number of data samples is partitioned into training, validation, and testing samples. For the three cases listed in Table \ref{tab:cases} the number of training, validation, and testing data samples are presented. Although the data set contains only $O(1000)$ reference particles, the number of grid points available for training the prediction of flow inside is far more.

Normalized perturbations of velocity and pressure fields ($\bu^{\prime},\ p^{\prime}$) defined below are the expected outputs from the SE(3)-CNN:

\begin{equation}
	\bu^{\prime} = \frac{\bu-\langle \bu \rangle}{\lvert \langle \bu \rangle \rvert} \, , \quad
	p^{\prime} = \frac{p-\langle p \rangle}{3\sigma_{p}}
\end{equation}
Here, $\sigma_{p}$ is an estimate of rms pressure obtained using the formula presented in [\onlinecite{siddani2020machine}]. 

\section{Overview of SE(3)-CNN}\label{se3_overview}
This work makes use of the framework \href{https://github.com/e3nn/e3nn}{e3nn} [\onlinecite{e3nn_code}, \onlinecite{e3nn_tutorial}] that is built on top of PyTorch [\onlinecite{pytorch}] to implement the SE(3)-CNN architecture used to produce results presented in this article. The reader is suggested to read \citet{weiler20183d,thomas2018tensor} and references therein for a comprehensive understanding of SE(3)-CNN. Here we will provide only the essential information needed for the present purposes.

Consider a feature vector field $q(\bx)$, where the feature vector may consist of several scalar, vector and tensor quantities. For example, if the feature is made up of Reynolds number, pressure, velocity and stress tensor, the length of the feature vector is 14, corresponding to 2 scalars, 1 vector and 1 second-rank tensor. Each element of the feature vector is a three-dimensional field (the Reynolds number is also defined as a constant field). We now define the set of all possible rotations about the origin to be the group $G$. Let us consider the action of a rotation $g \in G$ on the feature vector field. A post-rotation feature vector at point $\bx$, before rotation corresponds to a pre-rotation feature vector at a point $\by = \bR^{-1}(g) \, \bx$ , where $\bR(g)$ is a $3\times3$ rotation matrix corresponding to the rotational group element $g$. Furthermore, after rotation, (i) a scalar element of the feature vector, $p$ remains invariant to become $\rho_0(g) p$, where $\rho_0(g) = 1$ (ii) a vector $\bu$ transforms as $\rho_1(g) \, \bu$, where $\rho_1(g) = \bR(g)$, and (iii) a second rank tensor $\sigma$ transform as $\rho_2(g) \, \sigma$, where $\rho_2 = \bR(g) \, \bigotimes_{Kron} \,\bR(g)$ is a $9 \times 9$ matrix formed by the kronecker product, $\bigotimes_{Kron}$, of the rotational matrix with itself (note the second rank tensor has been unrolled into a 9-element long vector). Since, the rotational transformation of each scalar, vector and tensor component is independent the rotational transformation of the feature vector field can be expressed as
\begin{equation} \label{eq:pi-rho}
    \left[\pi(g) q\right] (\bx) = \rho(g) \, q(\bR^{-1}(g) \, \bx) \, 
\end{equation}
where $\rho$ corresponds to the rotational representation of $G$ for the entire feature vector, which can be expressed as
\begin{equation}
    \rho(g) = \bigoplus_{i=0}^{I} \bigoplus_{j=1}^{n_i} \rho_i(g) \, ,
\end{equation}
where $\bigoplus$ corresponds to construction of a block diagonal matrix with $n_i$ copies of the matrix operators $\rho_i$ for $i = 0$, $1, \cdots, I$. In the above notation, the feature vector $q$ contains $n_0$ scalars, $n_1$ vectors, $n_2$ second rank tensors and so on upto $n_I$ $I$th rank Cartesian tensors and therefore is of length $N=\sum_{i=0}^I n_i (3)^i$. In \eqref{eq:pi-rho}, $\pi(g)$ is the map that appropriately transforms the feature vector $q$ at the point $\by$ to $\rho(g)q$ at point $\bx$.

Now that rotational representations of arbitrary feature vectors have been presented, we can mathematically define the rotational equivariance of a CNN that takes in an input feature vector $q$ to produce the desired output feature vector $s$. A rotationally-equivariant CNN can be considered as a functional relation $F$ between the input and the desired output feature vectors that changes \textit{deterministically} under the operation of 3D rotational group elements. In the present problem, the input feature vector $q$ is a particular configuration of relative location of $P$ neighbors within the domain ($P$ vectors), particle Reynolds number (scalar), along with the unit vector along the average relative fluid velocity (vector). There are infinite possible neighbor configurations, particle Reynolds numbers and relative velocity orientations, which together define the set $Q$, and $q \in Q$. Let the flow field corresponding to the input feature $q$ be $s(\bx)$, and the set of flow fields of all input feature vectors be $S$. From this definition $s \in S$. Thus, the quest to predict the flow given an input feature vector can be expressed as a functional relation $F :\ Q \longrightarrow S $. The functional relation $F$ is said to be equivariant to an element $g \in G$ if:
\begin{equation}\label{eq:equivariance_eqn}
    F\left(\pi^{Q}(g) \, q \right) = \pi^{S}(g) \, F\left(q\right)
\end{equation}
where $\pi^{Q}: \ Q \longrightarrow Q$ is the set of rotational representations of $G$ on $Q$ and $\pi^{S}: S \longrightarrow S$ is the set of rotational representations of $G$ on $S$ [\onlinecite{satorras2021en}]. 

The above definition given in \eqref{eq:equivariance_eqn} relates the rotational equivariance of the relationship between the overall input and output features of the CNN. However, in a deep neural network, the input and output features are separated by many intermediate layers. It can be mathematically shown that for the overall CNN to be rotationally equivariant, each layer of the CNN must be equivariant. Several linear and non-linear layers put together form a constitutive block of a CNN. The following three steps are essential in order to guarantee the rotational equivaraiance of each block of CNN:
\begin{itemize}
    \item The input and output feature vectors of each block must be irreducible representations of the SO(3) group. Therefore, in the input feature vector of the CNN (which is the input to the very first layer) cartesian fields such as pressure, velocity, vorticity and stress/strain tensors must be transformed from raw quantities to irreducible representations. Similarly, the output feature vector of the CNN must be synthesized from the irreducible representation output of the last block.
    \item The convolution operation performed in each block must be a rotationally equivariant linear map. This will guarantee that the functional relationship between the input and output of the block satisfies rotational equivariance similar to that given in \eqref{eq:equivariance_eqn}.
    \item Non-linearities included in a block must also be rotationally equivariant, which implies that that non-linearities must be carefully applied to each irreducible representation rather than individual elements of the feature vector. 
\end{itemize}
Each of these aspects will be elaborated below.

\subsection{Irreducible features of SO(3)}
The SO(3) group contains all possible 3D rotations about the origin as group elements.  Tensor field physical properties, such as pressure, velocity and strain can be reduced further into representations known as \textit{irreducible representations} of SO(3). These irreducible representations cannot be decomposed further into a direct sum of representations.

A scalar quantity like pressure is an order-0 irreducible representation as it is invariant under rotational transformation. Vectors and pseudo-vectors, such as velocity and vorticity respectively, are order-1 irreducible representations as there is intertwining of the three Cartesian components when subjected to an arbitrary rotation, and therefore, cannot be broken down further. Cartesian tensors of rank greater than 1 can be decomposed into a summation of irreducible representations. Decomposition of a second rank Cartesian tensor, $\sigma$, has been shown below as an example.

\begin{equation}\label{eq:rank2_cart}
    \sigma = \underbrace{\left(\frac{tr(\sigma)}{3} \, \mathbf{I}\right)}_{\mbox{Order 0}} + \underbrace{\left( \frac{\sigma - \sigma^{T}}{2} \right)}_{\mbox{Order 1}} + \underbrace{\left(\frac{\sigma + \sigma^{T}}{2} - \frac{tr(\sigma)}{3} \, \mathbf{I}\right)}_{\mbox{Order 2}}
\end{equation}
where, $tr(\sigma)$ corresponds to the trace of $\sigma$. In \eqref{eq:rank2_cart}, $\sigma$ has been reduced into three different terms shown in the right hand side of the equation. The first term depends only on trace of $\sigma$ is a scalar (order-0 representation). The second term represents the anti-symmetric part of $\sigma$ is an order-1 irreducible representation. The traceless symmetric part of $\sigma$ (third term on the right hand side) has five independent elements and is an order-2 irreducible representation. Therefore, a second rank Cartesian tensor has been decomposed into a summation of independent irreducible representations of order-0, 1 and 2, which have 1, 3 and 5 elements respectively. In general, a rank $l$ Cartesian tensor can be reduced into a summation of irreducible representations of order-$l$ and lower. An irreducible representation of order-$l$, where $l = 0, 1, 2, ...$, has a dimension of $2l+1$, i.e., $2l+1$ elements. Upon applying this transformation to the irreducible form, the feature vector $\hat{q}$ can be rearranged in terms of concatenation of $J$ irreducible components as
\begin{equation}
    \hat{q} = \bigoplus_{j=0}^{J} \hat{q}_{j}
\end{equation}
where the $j$th component $\hat{q}_j$ is of order-$l(j)$ and thus its length is $(2l(j)+1)$. Therefore the total length of $\hat{q}$ is $\sum_{j=0}^J (2l(j)+1)$. 

Correspondingly, the rotational representation $\rho$ can also be written in terms of irreducible representations. Such irreducible rotational representations of order-0, 1 and 2 are
\begin{equation}
    \rho_0(g) = D^0(g) \, , \quad 
    \rho_1(g) = D^1(g) \, , \quad
    \rho_2(g) = T_2^{-1} \left[ \bigoplus_{m=0}^{2}D^m(g) \right] \, T_2 \, ,
\end{equation}
where $D^m(g)$ is the irreducible representation of $G$ at order-$m$. It is also known as the Wigner-D matrix of order-$m$ and it is of size $(2m+1)\times(2m+1)$. Thus, the $9\times9$ block representation matrix $\rho_2$ can be further decomposed to block representations of size $1\times1$, $3\times3$ and $5\times5$. Each of these irreducible rotational representations do not mix and therefore can be considered independently. Here $T_2$ is the change of basis matrix that decomposes a second rank tensor into its order-0, order-1 and order-2 irreducible representations. Also, $T_2^{-1}$ is the inverse which will assemble the order-0, order-1 and order-2 irreducible representations back into a second rank tensor. 

We note that the decomposition operator $T_i$ needs to be applied only once for the input feature vector of the entire CNN, and the resulting irreducible feature vector is the input to the first layer. After that all layer input and output features are in terms of irreducible feature vectors. Finally, the output of the last layer is assembled back with the change of basis $T_i^{-1}$ to obtain the desired output feature vector. Hence, all further discussion will assume the input and output feature vectors of each layer to be in irreducible form. 

It must be emphasized that the rotationally equivariant convolution operation of SE(3)-CNN, explained in subsection \ref{equivariant_kernel_subsection}, relies on the \textit{tensor product of irreducible representations} [\onlinecite{thomas2018tensor}] involving the convolution kernel at each block of the CNN. This explains the reason behind the input and output features vectors of each block being irreducible representations, since only with such an irreducible input feature vector the convolution operation can guarantee rotational equivariance.

\subsection{Tensor Product of Irreducible Representations}
Two irreducible representations of different orders can be combined together using `tensor product of irreducible representations', $\bigotimes_{Irreps}$, to produce new irreducible representations of different orders. Let us consider two irreducible representations, namely, $\hat{q}_{1}$ and $\hat{q}_{2}$ of orders $l_1$ and $l_2$ respectively. Then the `tensor product of irreducible representations' between $\hat{q}_1$ and $\hat{q}_2$ gives rise to irreducible representation $\hat{q}_3$ of order $l_3$ for $\abs{l_1-l_2} \leq l_3 \leq (l_1+l_2)$. The tensor product of each output irreducible representation will be denoted by $\left(\hat{q}_{1} \ \bigotimes_{Irreps}^{(l_1,l_2,l_3)} \ \hat{q}_{2} = \hat{q}_{3}\right)$ and given by
\begin{equation}\label{eq:tp_irreps}
    \hat{q}_{3,k}(\bx) = \sum_{i=1}^{(2l_1+1)} \ \sum_{j=1}^{(2l_2+1)} C_{i,j,k}^{(l_1,l_2,l_3)} \ \hat{q}_{1,i}(\bx) \ \hat{q}_{2,j}(\bx), \quad \text{for} \quad k = 1,2,...,2l_3+1
\end{equation}
where $C_{i,j,k}^{(l_1,l_2,l_3)}$ are (real) Clebsch-Gordan coefficients. The indices $i$ \& $j$ are along the dimensions of $\hat{q}_1$ \&  $\hat{q}_2$ respectively, and similarly, $k$ is along the length of $\hat{q}_3$. Furthermore, the `tensor product of irreducible representations' is equivariant [\onlinecite{thomas2018tensor}].

It can be seen from \eqref{eq:tp_irreps} that `tensor product of irreducible representations' can be broken down into two steps. The first step being outer product between $\hat{q}_{1}$ and $\hat{q}_{2}$, $\left(\hat{q}_{1} \ \bigotimes_{outer} \ \hat{q}_{2} = \hat{q}_{1,i} \ \hat{q}_{2,j}\right)$. The resulting $(2l_1+1) \times (2l_2+1)$ matrix is reshaped into a column matrix of size $(2l_1+1)(2l_2+1) \times 1$. Secondly, the Clebsch-Gordan coefficients can also be appropriately arranged into a matrix of size $(2l_3+1) \times (2l_1+1) (2l_2+1)$. Matrix multiplication between these two matrices generates $\hat{q}_{3}$ of dimension $2l_3+1$. Essentially, the Clebsch-Gordan coefficients form a change of basis matrix that is utilized in creating irreducible representation $\hat{q}_{3}$ from the outer product of $\hat{q}_{1}$ and $\hat{q}_{2}$. Note that this two-step operation can be carried out for only $\abs{l_1-l_2} \leq l_3 \leq (l_1+l_2)$. This two-step perspective is useful in appreciating similarity between conventional-convolution and equivariant-convolution operations discussed in the following subsections.    

\subsection{Convolution Kernel of a Conventional-CNN}
The convolution operation can be described as the most important mathematical operation in a CNN. The layer performing the convolution operation is known as a \textit{convolutional layer}. The convolution operations performed in a conventional-CNN and SE(3)-CNN will be described to highlight their differences. Again, in both cases, the convolution operation needs to be described for only one block of the neural network, since this process is repeated for all blocks of the CNN. 

In the conventional-CNN, the input feature vector, denoted as $q$, is a vector of length $N_{LI}$ consisting of stacked list of $n_{0LI}$ zeroth, $n_{1LI}$ first, and $n_{2LI}$ second rank tensors. Similarly, let the output feature vector of the layer, $s$, be made up of $n_{0LO}$ zeroth, $n_{1LO}$ first, and $n_{2LO}$ second rank tensors with a total length of $N_{LO}$ (here subscripts $LI$ and $LO$ stand for layer-input and layer-output). Let us also assume that a convolution operation is applied to map the input features to the output features. Then, the convolution operation performed in a conventional CNN is as follows.
\begin{equation}\label{eq:regular_cnn_conv}
    s_{i}(\bx_{1}) = \int\limits_{\mathbb{R}^3}  \sum_{j = 1}^{N_{LI}} \kappa_{ij}(\bx_2 - \bx_1) \ q_j (\bx_2) \, d\bx_2, \quad
    \text{for} \quad i = 1,2,...,N_{LO}
\end{equation}
Here, $\bx_{1}$ is the center of convolution operation, $\kappa_{ij} (\bx_{2}-\bx_{1})$ is the continuous learnable kernel, and $\bx_{2}$ is the dummy variable spanning over the 3D space ($\mathbb{R}^3$). The index $i$ runs along the dimension of the output feature vector $s$ and $j$ moves along the dimension of the input feature vector $q$. The learnable kernel takes relative position about $\bx_{1}$ as input and contains parameters that can be optimized so that $q$ can be linearly mapped to $s$. The aspect of considering relative location ensures that the CNN preserves translational invariance. The kernel is a matrix operator of size $N_{LO} \times N_{LI}$ and it is a component of the matrix multiplication ($\sum_{j} \kappa_{i,j} \ q_{j}$) of the convolution operation. The kernel can be recognized to have a block structure, where each block relates one of the zeroth, first, or second rank feature of the input vector with one of the zeroth, first, or second rank feature of the output vector. Thus, each block is of size $3^k \times 3^m$, where $k$ and $m$ are the rank of the output and input feature, respectively.

Though the convolution operation is shown in continuous space, during implementation the space is discretized and the information is only available at discrete grid points. Similarly, the size of the kernel along each direction is chosen so that only the points that fall within the kernel size about the center of convolution are considered and not the entire space.

\subsection{Equivariant Kernel of a SE(3)-CNN}\label{equivariant_kernel_subsection}
The conventional convolution operation given in \eqref{eq:regular_cnn_conv}, when considered as a functional relation between $q$ and $s$, does not satisfy the equivariance condition stated in \eqref{eq:equivariance_eqn}. Since the input and output feature vectors transform according to their representations, the convolution operation that maps the input features to the output features can also be expressed in terms of equivariant sub-kernel operations, when expressed in irreducible form. The equivariant sub-kernels used in the convolution operation are constrained to be built from (real) spherical harmonics (see [\onlinecite{weiler20183d}] for proof). Spherical harmonics are irreducible representations of the SO(3) group. Moreover, it is known that the `tensor product of irreducible representations' is an irreducible representation itself and it is equivariant as well [\onlinecite{thomas2018tensor}]. Thus, the decomposition of the input and output Cartesian fields into irreducible representations will now ensure that the convolution operation between them can be built using equivariant sub-kernels and tensor product of irreducible representations.

In the SE(3)-CNN, the input feature vector of a layer will be denoted as $\hat{q}$, and it is in irreducible form. It is also of length $N_{LI}$, but is made up of a stacked list of $J_{LI}$ irreducible components. Similarly, the output feature vector $\hat{s}$ is made up of $J_{LO}$ irreducible components, with a total length of $N_{LO}$. The SE(3)-equivariant convolution operation performed on $\hat{q}$ in a single layer of the network to obtain the output feature vector $\hat{s}$ can be expressed as
\begin{equation}\label{eq:se3cnn_conv}
    \hat{s}_k(\bx_1) = \int\limits_{\mathbb{R}^3} \sum_{j=1}^{J_{LI}} \ \eta_{k,j}(\bx_2-\bx_1)  \bigotimes\nolimits_{Irreps}^{(L,l_{LI}(j),l_{LO}(k))} \hat{q}_{j}(\bx_2)  \, d\bx_2, \quad \text{for} \quad k = 1,2,...,J_{LO}
\end{equation}
where $\hat{q}_j$ denotes the $j$th irreducible component of the input feature of order-$l_{LI}{(j)}$ and $\hat{s}_k$ corresponds to the $k$th irreducible component of the output feature of order-$l_{LO}{(k)}$. Here, $\eta_{k,j}$ represents the list of all irreducible representations that will preserve the equivariance between input feature, $\hat{q}_{j}$, and output feature, $\hat{s}_{k}$. Finally, the variable $L$ corresponds to order of each individual representation of $\eta_{k,j}$. The symbol $\bigotimes_{Irreps}^{(L,l_{LI}(j),l_{LO}(k))}$ denotes that each individual irreducible representation of $\eta_{k,j}$ tensor products with $\hat{q}_{j}$ to produce an irreducible representation of order-$l_{LO}(k)$.

The equivariance preserving list of irreducible representations, $\eta_{k,j}$, is given by \eqref{eq:se3_eta}.

\begin{equation}\label{eq:se3_eta}
    \eta_{k,j}(\bx_2-\bx_1) = \bigoplus_{L=\abs{l_{LO}(k)-l_{LI}(j)}}^{(l_{LO}(k)+l_{LI}(j))} \eta_{L,k,j}(\abs{\bx_2-\bx_1}) \ Y^{L}\left(\frac{\bx_2-\bx_1}{\abs{\bx_2-\bx_1}}\right)
\end{equation}
where, $\bigoplus$ denotes concatenation of the different irreducible representations and $Y^{L} = \left(Y_{-L}^{L},....,Y_{L}^{L}\right) \in \mathbb{R}^{2L+1}$ corresponds to (real) spherical harmonics of order $L$. Unlike the traditional convolution operation, the learnable part of $\eta_{k,j}$, given by $\eta_{L,k,j}$, is limited to take only the radial distance $(\abs{\bx_2-\bx_1})$ between the points $\bx_1$ and $\bx_2$ as input. The polar ($\theta$) and azimuthal ($\Phi$) (spherical coordinate system) dependence $\left(\text{i.e., dependence on }\frac{\bx_2-\bx_1}{\abs{\bx_2-\bx_1}}\right)$ is restricted to (real) spherical harmonics, where $Y^{L}(\theta,\Phi) \in \mathbb{R}^{2L+1}$ are (real) spherical harmonics of order $L$. The limits $\abs{l_{LO}(k)-l_{LI}(j)} \leq L \leq (l_{LO}(k)+l_{LI}(j))$ denote the range of spherical harmonics that can lead to an irreducible representation of order-$l_{LO}(k)$. 

The following steps will show how the different irreducible representations of $\eta_{k,j}$ can be combined together. The total length of $\eta_{k,j}$ is given by $\sum_{L=\abs{l_{LO}(k)-l_{LI}(j)}}^{(l_{LO}(k)+l_{LI}(j))} 2L+1 = (2l_{LO}(k)+1)(2l_{LI}(j)+1)$. It has been explained earlier that the Clebsch-Gordan coefficients involved in $\bigotimes_{Irreps}$ can be assembled into a change of basis matrix. Similarly, the coefficients involved in all the different $\bigotimes_{Irreps}$ between $\eta_{k,j}$ and $\hat{q}_{j}$ can also be arranged into a change of basis matrix $T_{CG}$, which has a size of $(2l_{LO}(k)+1) (2l_{LI}(j)+1) \times (2l_{LO}(k)+1) (2l_{LI}(j)+1)$. Matrix multiplication between $T_{CG}$ and $\eta_{k,j}$ will produce a column vector. The column vector can be rearranged into a matrix of size $(2l_{LO}(k)+1) \times (2l_{LI}(j)+1)$, which can be denoted as $\hat{\kappa}_{k,j}$. Similar to the block form that exists in a conventional-convolution operation this sub-kernel block $\hat{\kappa}_{k,j}$ acts between the $j$th input irreducible feature and $k$th output irreducible feature. This also explains the size of the block being $(2l_{LO}(k)+1) \times (2l_{LI}(j)+1)$ as the output feature is of size $(2l_{LO}(k)+1)$ and the input feature is of size $(2l_{LI}(j)+1)$. 

\subsection{Equivariant Non-Linearity}
In traditional neural networks, the non-linearity using an activation function ($\mathcal{F}$) is applied elementwise to all of its input features. However, such a non-linearity cannot be applied in SE(3)-CNN as it compromises the equivariance property. An activation function {can} be applied elementwise for a {scalar} feature but {cannot} be applied elementwise for a {non-scalar} irreducible feature of SO(3) as it ignores the intertwining of the elements of an irreducible component. Thus, non-linearity is applied with the following two equivariant methodologies for non-scalar irreducible features.

\subsubsection{Norm-based activation}
The norm of a non-scalar irreducible feature, $\hat{s}_k$, of order-$l_{LO}(k)$ is given by
\begin{equation}
    \norm{\hat{s}_k} = \sqrt{\sum_{i=1}^{2l_{LO}(k)+1} \ \left(\hat{s}_{k,i}\right )^{2}} \, ,
\end{equation}
which does not change when $\hat{s}_k$ is subjected to a rotation or translation. As the norm is a scalar, the non-linearity is applied to the norm and then this value is multiplied with the irreducible feature, $\hat{s}_k$. Multiplying any irreducible feature with a scalar is equivariant. This activation [\onlinecite{thomas2018tensor}] is summarized as \eqref{eq:norm_act}. 
\begin{equation}\label{eq:norm_act}
    \text{Norm-based activation} =
    \begin{cases}
        \mathcal{F}(\hat{s}_k) & \text{if $\hat{s}_k$ is a scalar irreducible feature.} \\
        \mathcal{F}(\norm{\hat{s}_k}) \hat{s}_k & \text{if $\hat{s}_k$ is a non-scalar irreducible feature.}
    \end{cases}
\end{equation}

\subsubsection{Gated Non-Linearity}
The gated activation [\onlinecite{weiler20183d}] is also built on the same principle that multiplying any irreducible feature with a scalar is equivariant. Generally, non-linearity in CNNs is applied after a convolution operation. This method takes advantage of the previous convolution operation to produce additional scalar fields termed \textit{gates}. As these gates are produced using SE(3)-convolution they are equivariant. The non-linearity is applied to these scalar gates and then multiplied with non-scalar irreducible features. Thus, the number of gates produced is equal to the number of non-scalar irreducible features.

Let us consider that $J_{sLO}$ scalar features, \{$\hat{s}_k, \ k=1,2,...,J_{sLO} $\}, $J_{LO}-J_{sLO}$ non-scalar irreducible features, \{$\hat{s}_k, \ k=J_{sLO}+1,\cdots,J_{LO}$\}, and $J_{LO}-J_{sLO}$ scalar gates, \{$\gamma_k, \ k=J_{sLO}+1,\cdots,J_{LO}$\}, are produced as output features of a SE(3)-convolution process. Then the gated activation is applied as follows
\begin{equation}\label{eq:gate_act}
    \text{Gated activation} =  
    \begin{cases}
        \mathcal{F}(\hat{s}_k) & k = 1,2,\cdots,J_{sLO} \\
        \mathcal{F}(\gamma_k) \hat{s}_k & k=J_{sLO}+1,\cdots,J_{LO} 
    \end{cases}
\end{equation}
Hence, after the application of the gated non-linearity the total number of features is reduced back to $J_{LO}$.

\subsection{Constitutive Non-linear SE(3)-CNN block}
\begin{figure}[h]
    \centering
    \includegraphics[width=0.8\linewidth,keepaspectratio=true]{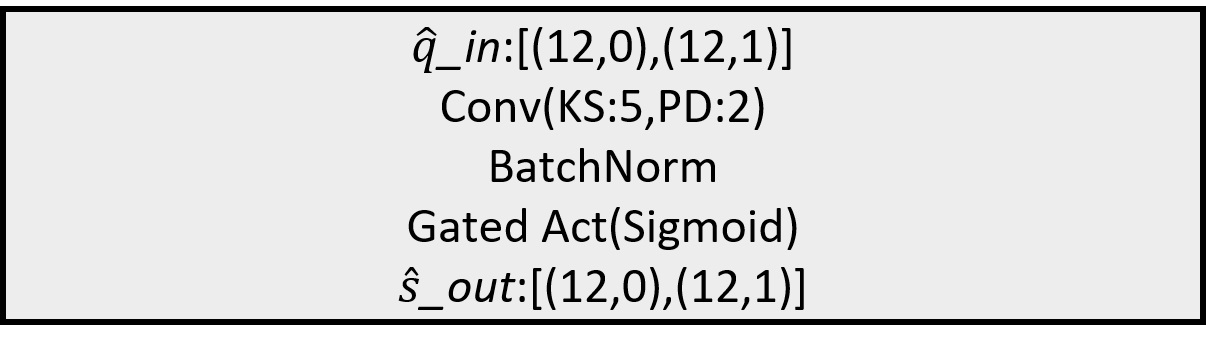}
    \caption{A non-linear SE(3)-CNN block.}
    \label{fig:se3_block}
\end{figure}
One block of the SE(3)-CNN is symbolically shown in Figure \ref{fig:se3_block}. The irreducible input feature vector to the layer is denoted by $\hat{q}\_in$\footnote{$\hat{q}\_in$ is the same as Rs\_in used in Figure \ref{fig:network} and the remainder of this paper.}. In this particular block, the (12,0) in $\hat{q}\_in$ corresponds to 12 different order-0 and the (12,1) corresponds to 12 different irreducible order-1 features. The final output features of the block are given by $\hat{s}\_out$\footnote{$\hat{s}\_out$ is the same as Rs\_out used in Figure \ref{fig:network} and the remainder of this article.}. The example shown in the figure produces 12 order-0 and 12 order-1 irreducible features as outputs. The mathematical operations performed in this block are convolution operation (\textbf{Conv}), batch normalization (\textbf{BatchNorm}) and gated activation (\textbf{Gated Act}). The sequence of these operations is same as the order in which they appear from top to bottom.

As $\hat{s}\_out$ is the final output produced after the gated non-linearity, irreducible representations produced by the earlier applied convolution operation are 12 scalars, 12 scalar gates and 12 order-1 irreducible features. The convolution operation has a kernel of size 5 (\textbf{KS}) along each of the three directions. The zero padding (\textbf{PD}) used along each boundary of the domain is two in this example shown in the figure. The batch normalization used in SE(3)-CNN is an equivariant version of the regular batch normalization [\onlinecite{batch_norm}]. The reader is directed to [\onlinecite{e3nn_code}] for the implementation of the equivariant batch normalization. Finally, the predefined function used in the gated non-linearity is sigmoid function.
 
\section{ML Approach} \label{MLapp}
\begin{figure}
	\centering
	\includegraphics[width=\linewidth,keepaspectratio=true]{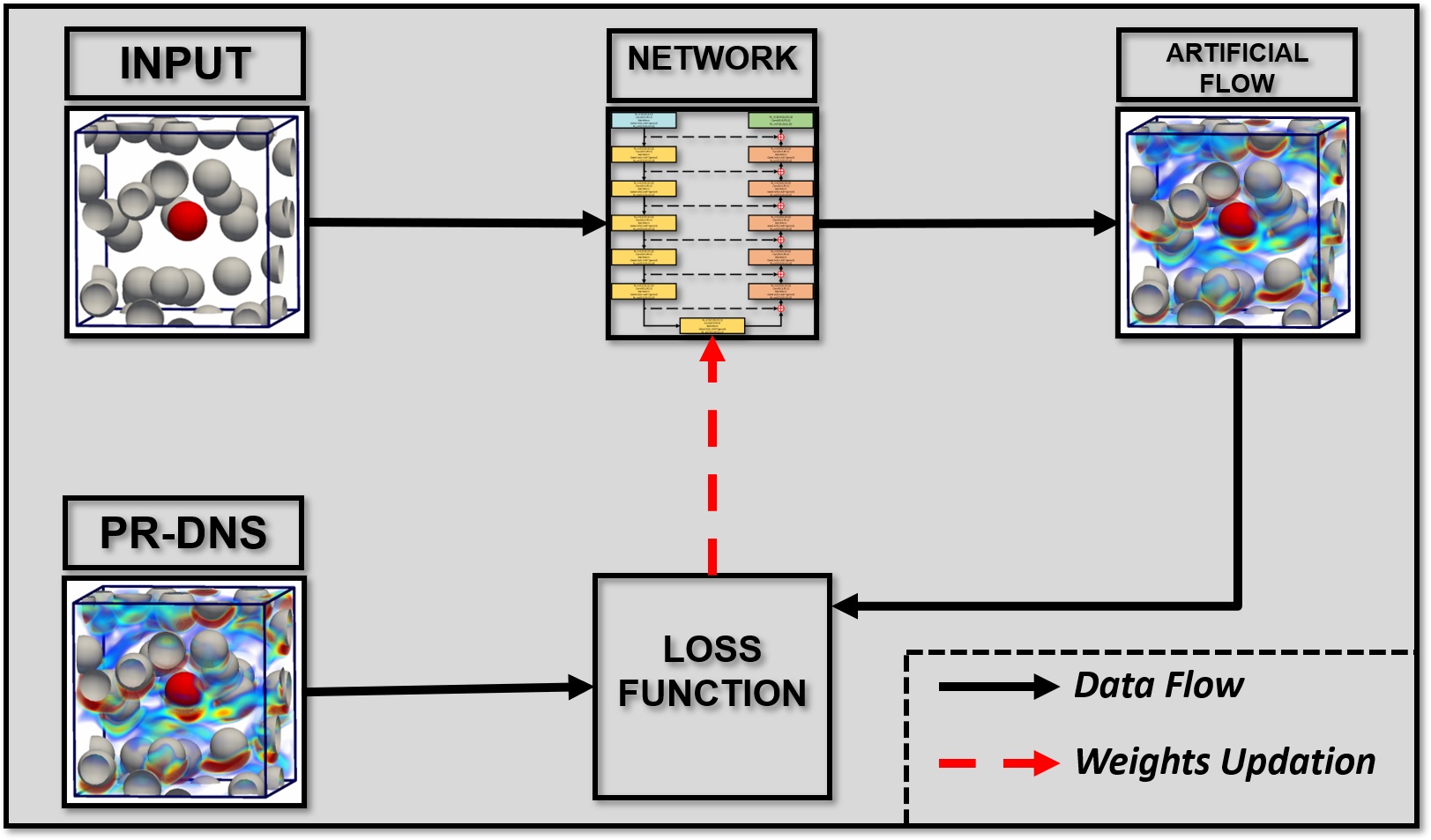}
	\caption{\textit{ML Approach}.}
	\label{fig:ML_approach}
\end{figure}

Pictorial representation of the overall ML approach implemented in this work is shown in Figure~\ref{fig:ML_approach}. As seen in the figure, artificial flow fields generated by the network are evaluated against the corresponding PR-DNS results using a loss function to improve the network's ability to recreate the PR-DNS flow fields.

64 uniformly spaced grid points are chosen along each direction in a sample. 
In this study, we will only concentrate on achieving accurate flow fields on a moderately coarse grid which takes all neighbors that fall in $5 \times 5 \times 5$ local volume into consideration. The input information contains locations of all particles (including the reference particle), volume-averaged Reynolds number and the mean flow direction. Discrete representation of locations of all particles in a sample can be achieved using Indicator Function ($I_{f}$), which  is defined as follows.
\begin{equation}
I_f(x,y,z) = \left\{
\begin{array}{ll}
0, & if\;(x,y,z)\;is\;inside\; a\;particle.\\[2pt]
1, & if\;(x,y,z)\;is\;outside\; a\;particle.
\end{array} \right.
\end{equation}
In addition to the indicator function, another scalar input information is Reynolds number ($Re$). Thus, these two scalars can be combined together to create the following redefined indicator function ($c_{1}$)
\begin{equation} \label{re_if}
c_1(x,y,z) = \left\{
\begin{array}{ll}
0, & if\;(x,y,z)\;is\;inside\; a\;particle.\\[2pt]
Re, & if\;(x,y,z)\;is\;outside\; a\;particle.
\end{array} \right.
\end{equation} 
Finally, the information of mean flow direction is fed to the network using a unit vector $\widehat{\langle \bu \rangle}$ along the respective direction. The streamwise velocity direction is required to fully-define the problem as it enforces rotational symmetry only about the mean-flow direction.  
In essence, the input feature ($\bc$) to the network is a combination of a scalar and a vector as shown below
\begin{equation} \label{in_con}
	\bc = \{ c_1, \, \widehat{\langle \bu \rangle}\} \, .
\end{equation}  

\subsection{Performance Evaluation Metric}
The metric used in this study to evaluate the performance of a network is coefficient of determination ($R^2$). For example, the performance of the network for streamwise velocity component, $u^{\prime}$, will be defined as follows.
\begin{equation}
R^{2}_{u^{\prime}} = 1-\dfrac{\sum\limits_{i=1}^{N_{te}}\sum\limits_{volume}(u_{DNS}^{\prime}-u_{network}^{\prime})^{2}I_{f}}{\sum\limits_{i=1}^{N_{te}}\sum\limits_{volume}(u_{DNS}^{\prime})^{2}I_{f}}
\end{equation}
where, $N_{te}$ is the number of test samples. Similarly, we can also evaluate the performance of the network for pressure ($p^{\prime}$) and transverse velocities ($v^{\prime}, \, w^{\prime}$).

$R^2$ value of unity denotes a perfect model that can exactly capture the PR-DNS fluctuations. A value of zero for a network is equivalent to not predicting any fluctuations at all. Finally, a value less than zero is an indication that the network is severely under-predicting or over-predicting the perturbations.
    
\subsection{Network, Loss Function, and Optimizer}
\begin{figure}
	\centering
	\includegraphics[width=\linewidth,keepaspectratio=true]{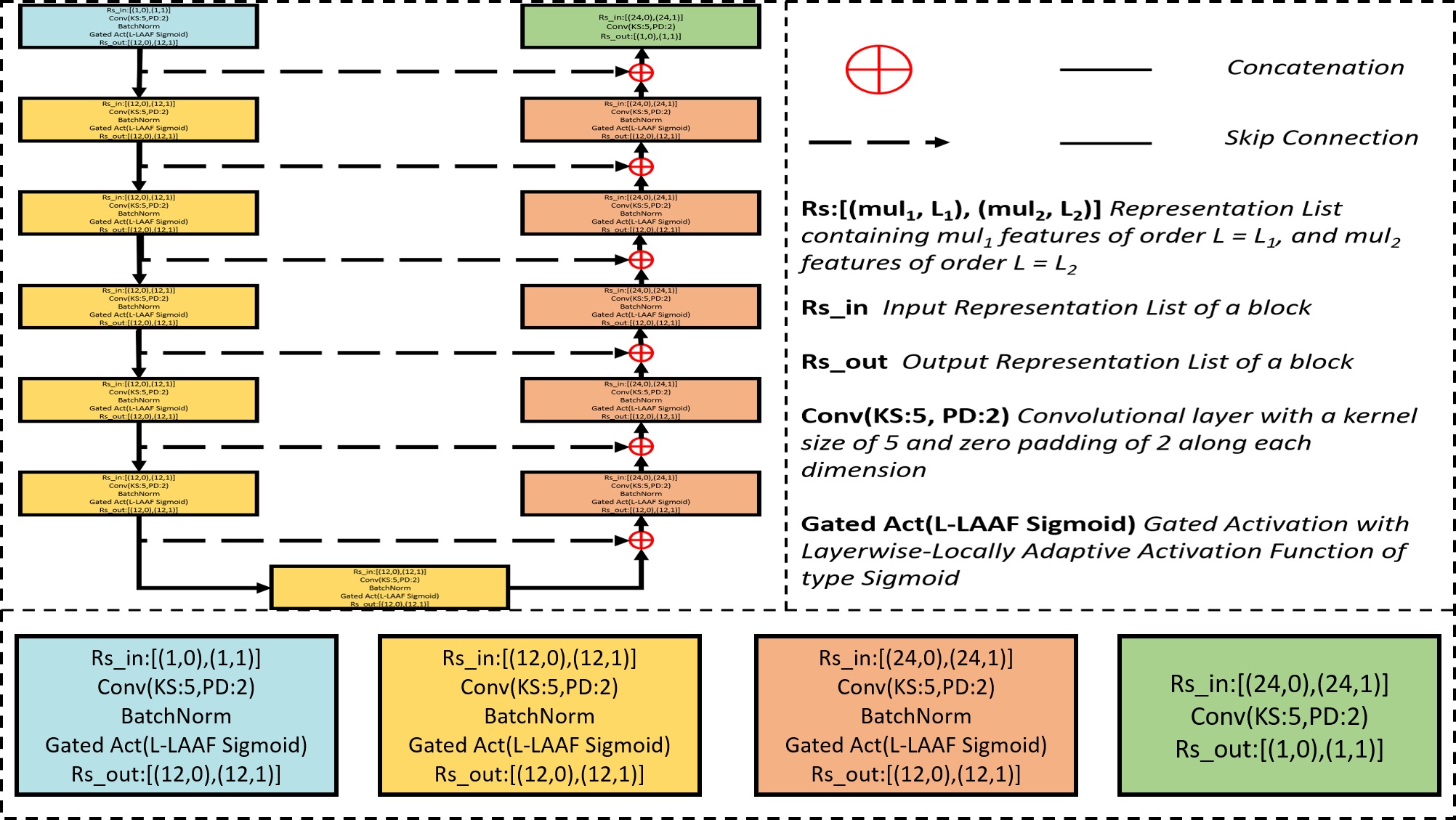}
	\caption{\textit{Network Architecture}.}
	\label{fig:network}
\end{figure}

Deep learning results presented in this work have been performed using a single Quadro RTX 6000 GPU device. The largest equivariant U-Net [\onlinecite{unet}] that could be considered for a nominal mini-batch of size 3 on this GPU device is a 13 block U-Net. The architecture of the network has been shown in Figure \ref{fig:network}. In the first block, the input information $\bc$ is the \textbf{Rs\_in}, i.e., one scalar and one vector [(1, 0), (1, 1)]. A filter/kernel of size (\textbf{KS} = 5) along each dimension has been used in every equivariant convolution operation (\textbf{Conv}) performed in this network. A zero-padding of size (\textbf{PD} = 2) for each boundary has been used to make sure that a $64 \times 64 \times 64$ output is obtained after applying a convolution operation. An equivariant Batch Normalization (\textbf{BatchNorm}) (see [\onlinecite{e3nn_code}]) is applied after the convolution process in the first 12 blocks. The final operation in a block after \textbf{BatchNorm} is the application of non-linearity. Here, gated non-linearity (see \eqref{eq:gate_act} and [\onlinecite{weiler20183d}]) has been utilized. The inputs to the network and its outputs are only scalars and vectors. Hence, the outputs of intermediate blocks (\textbf{Rs\_out}) have also been restricted to only representations of order-0 and 1. Each intermediate block outputs 12 scalars and 12 order-1 irreducible feature after the application of the gated non-linearity, \textbf{Rs\_out} = [(12, 0), (12, 1)]. As a U-Net involves skip connections, the outputs of the first 6 blocks are also concatenated along the feature dimension as shown in the figure. This is responsible for the \textbf{Rs\_in} = [(24, 0), (24, 1)] in the last 6 blocks inspite of their previous blocks only outputting 12 each of order-0 and 1 representations.  

Layer-wise locally adaptive activation function (L-LAAF) [\onlinecite{adpt_actfun}] of type Sigmoid with slope recovery term has been used in the gated non-linearity with an intention to accelerate the training process. The pre-defined scaling factor, $n$, of L-LAAF was chosen to be unity, which is its lower bound ($n \geq 1$). Thus, the scalable parameter, $a$, of each layer is initialized with $a = 1$ as $n \times a = 1$ was the initialization condition used in [\onlinecite{adpt_actfun}]. 

The loss function ($\mathcal{L}$) used in network training is given below.
\begin{equation}\label{loss}
    \mathcal{L} = \mathcal{DL} + \frac{\lambda}{64^3} \times \mathcal{PL} + 10 \times \mathcal{S}(a) \quad ,
\end{equation}
where Data-based Loss ($\mathcal{DL}$) is\\
\begin{equation}
    \mathcal{DL} = \norm{[\{ \bu^{\prime}, p^{\prime} \}_{DNS} - \{ \bu^{\prime}, p^{\prime} \}_{network}] \odot I_f}_{1} \, .
\end{equation}
The operation $\odot$ corresponds to elementwise multiplication between every field variable and indicator function. It ensures that the loss function only takes grid points that fall inside the fluid volume into consideration. Here $\norm{.}_1$ denotes L-1 norm averaged by the number of elements.  

Physics-based Loss ($\mathcal{PL}$) is defined as
\begin{equation}
    \mathcal{PL} = \norm{\sum_{volume}{\{\bu^{\prime}, p^{\prime}\}_{network}\odot I_f}}_1 \, .
\end{equation}
based on the principle that average of perturbation must be zero. Although the governing equations can also be included in the $\mathcal{PL}$ term, they have been avoided here due to the additional computational cost that is associated with imposing them. However, it has to be stated that inclusion of governing equations has resulted in more generalizable models in other fluid flow problems [\onlinecite{jiang2020meshfreeflownet}]. 

The slope recovery term ($\mathcal{S}(a)$) is\\
\begin{equation}\label{slope_recovery}
    \mathcal{S}(a) = \frac{1}{\frac{1}{12} \, \sum_{k=1}^{12} exp(a^k)} \quad .
\end{equation}
As mentioned earlier, the slope recovery term, which is associated with L-LAAF, is added for accelerated training convergence. In the 13-block network, 12 instances of this activation are applied in the first 12 blocks and hence $a^k$ in \eqref{slope_recovery} corresponds to the scalable parameter of each layer.

An analysis optimizing the value of $\lambda$ has been carried out using the case-1 dataset as the Reynolds number associated with this set is the highest among the considered cases. The test performance achieved with different $\lambda$ values has been presented in Table \ref{tab:lambda_variation}. It can be seen from the results that the inclusion of this particular physics-based loss term in the cost/loss function does not lead to a significant and consistent higher performance. This is primarily due to the weak nature of this condition. Unlike imposing a governing equation which has to be satisfied at every grid point, this more flexible global condition deals with the overall volumetric output produced by the network. The training for remaining datasets presented in Table \ref{tab:cases} is performed using $\lambda = 0.01$ as it produces the best performance for pressure and second best performance for velocity components among the considered values. The optimizer used in this study is ADAM optimizer [\onlinecite{adam}] and a learning rate of 0.001 was used for the equivariant network.

\begin{table}[]
    \centering
    \begin{ruledtabular}
         \begin{tabular}{cccc}
            $\lambda$ & Pressure & Streamwise Velocity & Transverse Velocities \\
            \hline
            0 & 0.7487 & 0.6844 & \textbf{0.4960} \\
            0.001 & 0.7433 & 0.6851 & 0.4906 \\
            \textbf{0.01} & \textbf{0.7533} & 0.6865 & 0.4945\\
            0.1 & 0.7481 & \textbf{0.6889} & 0.4896 \\
            1 & 0.7479 & 0.6802 & 0.4863
         \end{tabular}
    \end{ruledtabular}
    \caption{Test performance ($R^2$) for different $\lambda$ of the network trained using case $Re = 172.96$, $\phi = 0.11$.}
    \label{tab:lambda_variation}
\end{table}

\section{Results and Discussion}
Deep learning analysis with the network was individually performed for each of the considered cases. The following procedure was adopted to divide the samples available in each case into \textit{training}, \textit{validation}, and \textit{testing} datasets. As every case has multiple realizations, two realizations are selected randomly with one of them being assigned as the validation set and the other being the testing set. The remaining realizations put together form the training set. Test performance variability study performed in [\onlinecite{siddani2020machine}] for the same cases indicates that the variation in test performance with respect to which realization is taken as testing data is very low. Hence, the results presented here will be a good estimate of the overall performance of the network.

\subsection{Performance of Equivariant CNN}
Here, the performance of the SE(3)-CNN will be evaluated and it will also be compared with that of an equivalent regular CNN with and without data augmentation. The regular CNN is obtained by replacing all the equivariant components of the network shown in Figure \ref{fig:network} with conventional non-equivariant components directly from PyTorch [\onlinecite{pytorch}]. Essentially, the gated non-linearity is replaced by conventional elementwise activation function of the same type (L-LAAF Sigmoid). Every SE(3)-convolutional layer in the intermediate blocks of SE(3)-CNN produces 12 order-0 features, 12 scalar gates and 12 order-1 features as its outputs. Therefore, each conventional-convolutional layer of regular CNN outputs 60 channels to maintain the same overall dimension. As elementwise non-linearity is applied in the regular CNN the output of each block also has 60 channels. This regular CNN has approximately 7.23 times the independent parameters in the equivariant version.

Data augmentation has been utilized for both equivariant and regular CNN. As SE(3)-CNN only imposes 3D rotational and translation symmetries, reflection about either $y$ or $z$ axis has been performed for all datasets. The data augmentations performed for the regular CNN are discrete $90^{0}$ rotations about the streamwise direction (along $x$) and reflections about $y \, \& \, z$ axes. This regular CNN with data augmented samples will be referred to as \textit{Data-augmented CNN} and the traditional CNN without any data augmentation will be called \textit{Simple CNN}. Therefore, the total number of samples for the Data-augmented CNN is 8 times those of Simple CNN and the total number of samples for SE(3)-CNN is twice the number of samples in Simple CNN. The same loss function, given in \eqref{loss} with $\lambda = 0.01$, and optimizer are used for all the networks. A learning rate of 0.0002 was used for Data-augmented CNN and Simple CNN due to the significantly higher number of independent parameters in them.
\begin{table}
    \centering
    \begin{ruledtabular}
         \begin{tabular}{cccccccccc}
         Case & \multicolumn{3}{c}{SE(3)-CNN} & \multicolumn{3}{c}{Data-augmented CNN} & \multicolumn{3}{c}{Simple CNN} \\
          & $p^{\prime}$ & $u^{\prime}$ & $v^{\prime}, \, w^{\prime}$ & $p^{\prime}$ & $u^{\prime}$ & $v^{\prime}, \, w^{\prime}$ & $p^{\prime}$ & $u^{\prime}$ & $v^{\prime}, \, w^{\prime}$ \\
          \hline
          1 & 0.7533 & 0.6865 & 0.4945 & 0.7128 & 0.6607 & 0.4628 & 0.7195 & 0.5933 & 0.2972\\
          2 & 0.7803 & 0.8076 & 0.7166 & 0.7797 & 0.7729 & 0.6838 & 0.7058 & 0.7375 & 0.4933\\
          3 & 0.5941 & 0.8297 & 0.6988 & 0.4932 & 0.8081& 0.6549 & 0.4275 & 0.7691 & 0.5712
         \end{tabular}
    \end{ruledtabular}
    \caption{Test performance ($R^2$) of SE(3)-CNN, Data-augmented CNN and Simple CNN.}
    \label{tab:domain_performance}
\end{table}

It can be seen from Table \ref{tab:domain_performance} that the SE(3)-CNN outperforms both Data-augmented CNN and Simple CNN for all cases. This improvement can be attributed to the fact that continuous rotational symmetry about the given input streamwise direction is imposed when an equivariant CNN is used to generate the flow fields. This process can be interpreted as providing continuous data augmentations of infinitesimal angle rotations about the given streamwise direction and it is known that a neural network becomes more generalizable when it is provided with a larger size of training samples. Hence, the higher test performance for pressure ($p^{\prime}$) and streamwise velocity component ($u^{\prime}$) of SE(3)-CNN can be attributed to the aspect that the abstract information learned in its intermediate layers is equivariant, and therefore, more generalizable. The significant improvement for transverse velocity components ($v^{\prime}, \ w^{\prime}$) in Data-augmented CNN when compared with Simple CNN is indicative that discrete rotations and reflections are very helpful in teaching the CNN that $v^{\prime}$ and $w^{\prime}$ are statistically the same. Much higher performance of SE(3)-CNN for transverse velocity components suggests that the SE(3)-CNN does a better job in understanding that these two components are truly the same in a statistical sense.     

The obvious advantage of SE(3)-CNN, apart from higher test performance, is that it can be readily used in test samples whose streamwise direction substantially deviates from the $+ x$ axis due to its equivariance property. On the contrary, the regular CNN should either be trained through data augmentation along these directions, which results in increase of training time, or coordinate transformation has to be performed on the testing samples to reorient the streamwise direction along $+ x$ axis and then later followed by another coordinate transformation to shift back to the original coordinate system of testing data.

\subsection{Effect of training dataset size}
The amount of training data that is typically available in fluid flow applications is low. This increases the importance of using rotational and reflectional symmetries for increasing amount of data available for training. In this section we will illustrate that even with limited amount of training data SE(3)-CNN is able to perform quite well as a result of its in-built symmetry enforcement which tends to greatly enhance the available data. The impact of training dataset size is evaluated using the case-2 dataset. The networks are trained with limited sub-samples of the entire training data and the results are shown in Table \ref{tab:training_size_effect}. These sub-samples are chosen randomly from the entire training dataset. The test performance ($R^{2}$) for the three networks trained using this reduced training data is presented in Table \ref{tab:training_size_effect} along with the test performance of total training samples for the purpose of comparison. 

It can be seen from the results that both Data-augmented CNN and Simple CNN perform inconsistently when the number of unique samples is small. The poor performance can be attributed to the large number of trainable parameters involved in them that would generally lead to overfitting, which causes low generalization. However, the SE(3)-CNN performs moderately even with very few samples which is an indication to its high generalizability. When the unique training samples are of $O(10)$, both SE(3)-CNN and Data-augmented CNN outperform Simple CNN by a significant margin, especially for transverse velocities. This is to be expected as SE(3)-CNN implicitly enforces continuous 3D rotational symmetry and the Data-augmented CNN has eight times more samples from data augmentation process. Furthermore, the difference in performance for transverse velocities between SE(3)-CNN and Data-augmented CNN is considerably large even when the training samples are of $O(10)$. This higher performance is due to the continuous rotational symmetry of SE(3)-CNN architecture.

From the earlier discussions it can be stated that SE(3)-CNN is an efficient and effective alternative to performing discrete rotation data augmentation procedure, which becomes a necessity in the limit of scarce training data. Hence, SE(3)-CNN is the ideal option to obtain a generalizable model when the training data is limited.

\begin{table}[]
    \centering
    \begin{ruledtabular}
    \begin{tabular}{*{10}{c}}
     Unique & \multicolumn{3}{c}{SE(3)-CNN} & \multicolumn{3}{c}{Data-augmented CNN } & \multicolumn{3}{c}{Simple CNN}\\
     training samples & $p^{\prime}$ & $u^{\prime}$ & $v^{\prime}, \, w^{\prime}$ & $p^{\prime}$ & $u^{\prime}$ & $v^{\prime}, \, w^{\prime}$ & $p^{\prime}$ & $u^{\prime}$ & $v^{\prime}, \, w^{\prime}$ \\
     \hline
     1 & 0.5354 & 0.6355 & 0.4906 & 0.3267 & 0.4921 & -0.1493 & 0.2516 & 0.3596 & -0.2843 \\
     5 & 0.6610 & 0.7255 & 0.5779 & 0.6465 & 0.6616 & 0.3202 & 0.2531 & 0.2185 & -0.0666 \\
     20 & 0.7380 & 0.7754 & 0.6546 & 0.6630 & 0.7396 & 0.5366 & 0.5892 & 0.6616 & 0.1756 \\
     75 & 0.7649 & 0.8000 & 0.7018 & 0.7402 & 0.7858 & 0.6360 & 0.6554 & 0.7112 & 0.4102 \\
     450 & 0.7803 & 0.8076 & 0.7166 & 0.7797 & 0.7729 & 0.6838 & 0.7058 & 0.7375 & 0.4933
    \end{tabular}
    \end{ruledtabular}
    \caption{Test performance ($R^2$) of SE(3)-CNN, Data-augmented CNN and Simple CNN when trained on limited samples of case-2.}
    \label{tab:training_size_effect}
\end{table}

\subsection{Illustration of SE(3)-CNN equivariance}
\begin{figure}[h]
    \centering
    \begin{subfigure}[b]{\textwidth}
        \centering
        \includegraphics[width=0.24\textwidth,keepaspectratio=true]{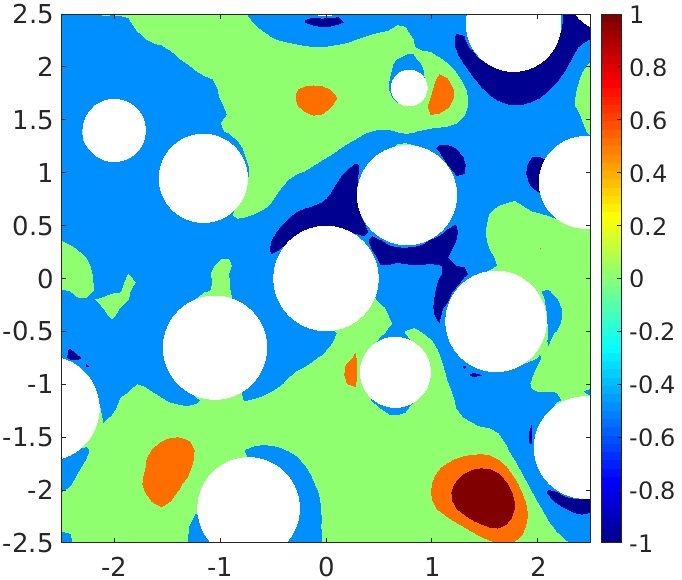}%
        \includegraphics[width=0.24\textwidth,keepaspectratio=true]{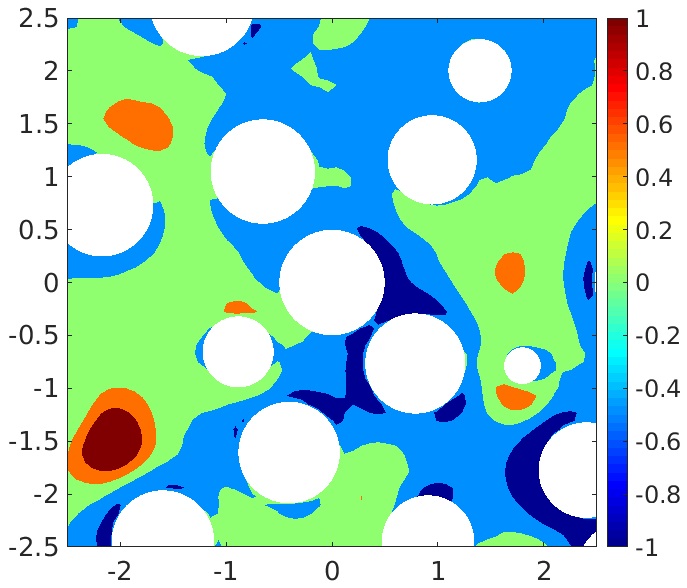}%
        \includegraphics[width=0.24\textwidth,keepaspectratio=true]{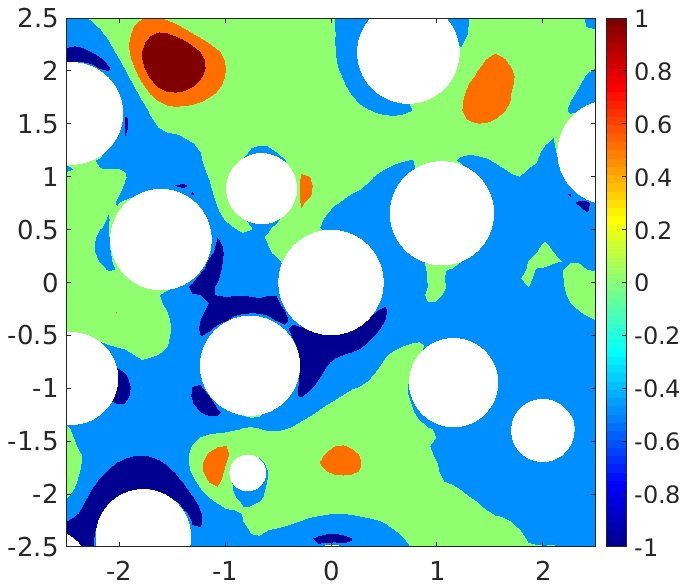}%
        \includegraphics[width=0.24\textwidth,keepaspectratio=true]{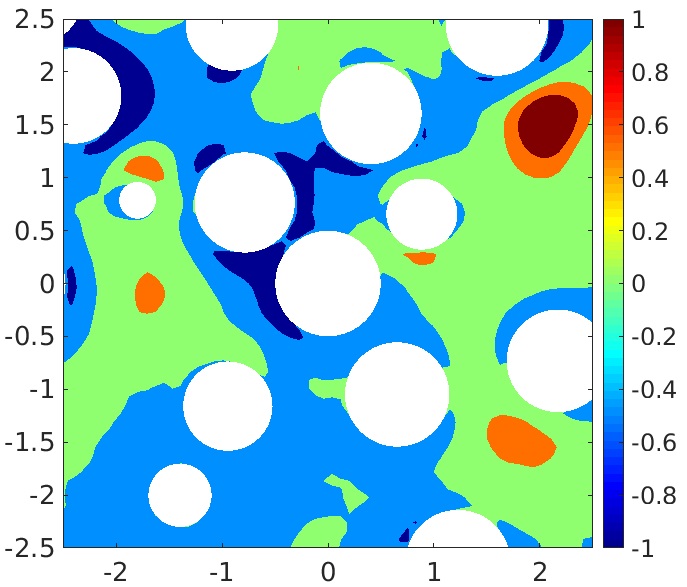}%
        \caption{SE(3)-CNN}
    \end{subfigure}
    \begin{subfigure}[b]{\textwidth}
        \centering
        \includegraphics[width=0.24\textwidth,keepaspectratio=true]{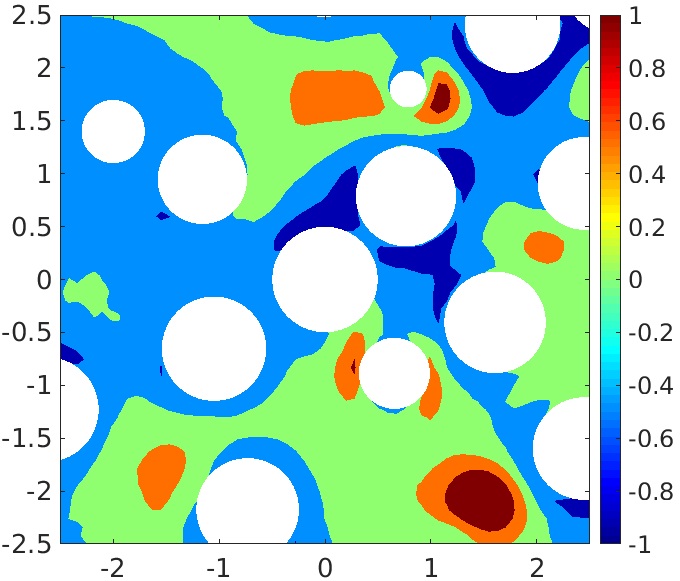}%
        \includegraphics[width=0.24\textwidth,keepaspectratio=true]{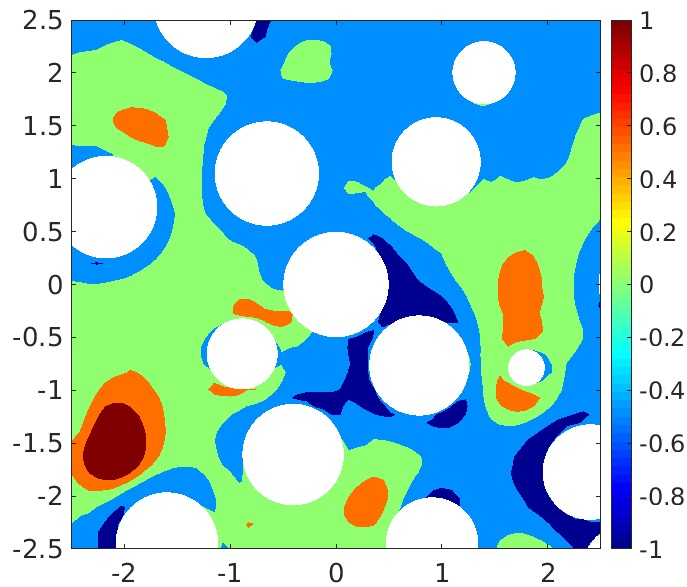}%
        \includegraphics[width=0.24\textwidth,keepaspectratio=true]{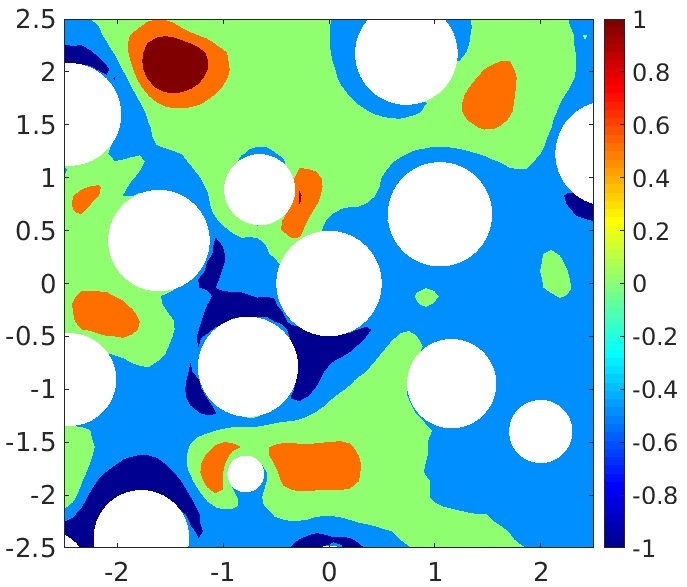}%
        \includegraphics[width=0.24\textwidth,keepaspectratio=true]{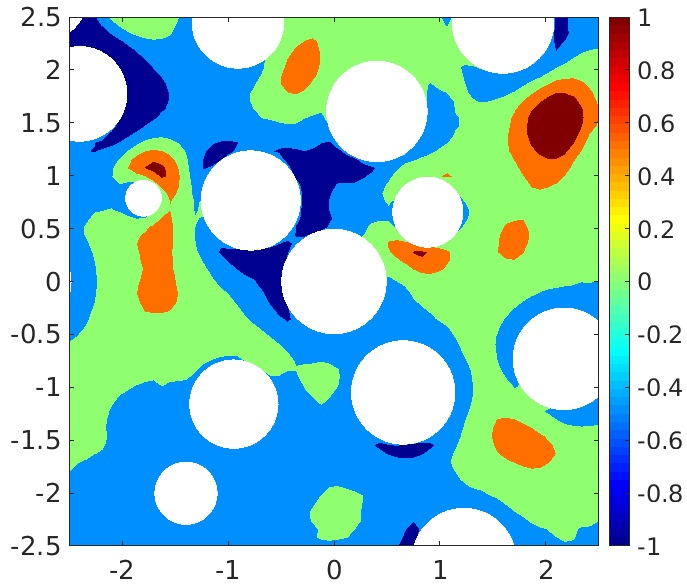}%
        \caption{Data-augmented CNN}
    \end{subfigure}
    \begin{subfigure}[b]{\textwidth}
        \centering
        \includegraphics[width=0.24\textwidth,keepaspectratio=true]{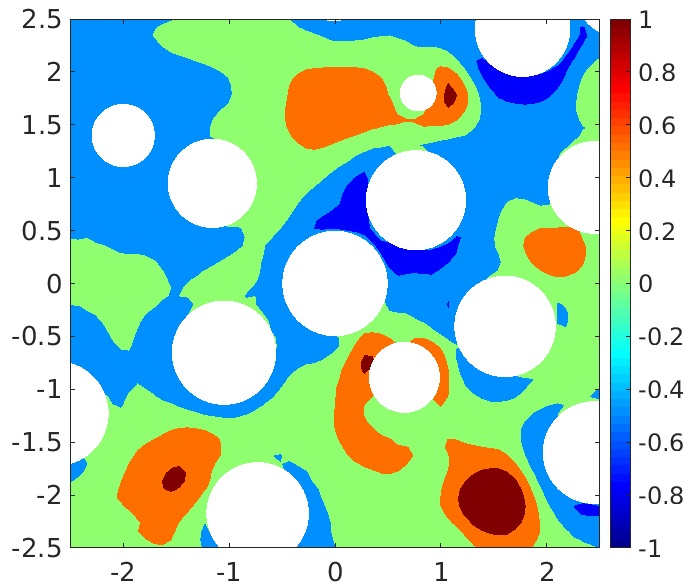}%
        \includegraphics[width=0.245\textwidth,keepaspectratio=true]{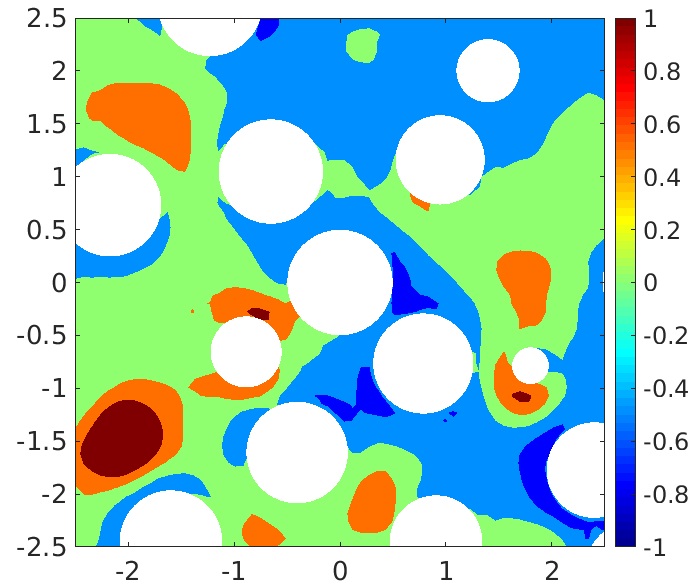}%
        \includegraphics[width=0.24\textwidth,keepaspectratio=true]{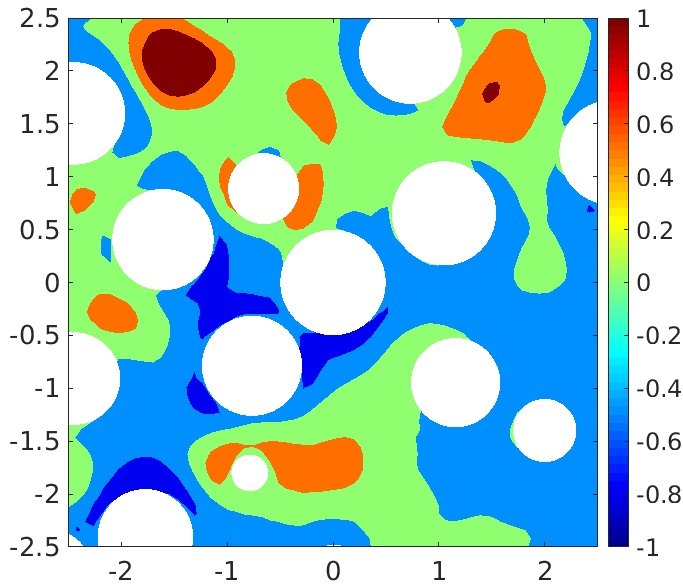}%
        \includegraphics[width=0.24\textwidth,keepaspectratio=true]{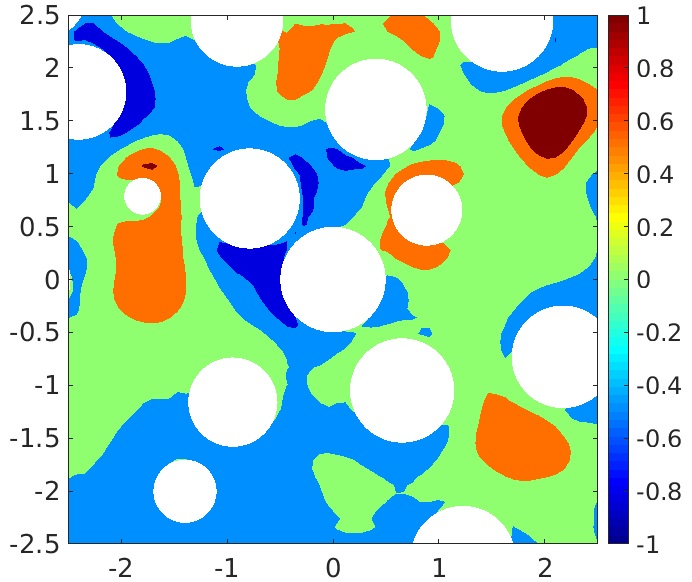}%
        \caption{Simple CNN}
    \end{subfigure}
    \caption{Pressure contours for a central $y-z$ plane of a test sample and its discrete rotations generated using (a) SE(3)-CNN, (b) Data-augmented CNN and (c) Simple CNN from case $Re = 86.22$, $\phi = 0.21$.}
    \label{fig:p_images}
\end{figure}

\begin{figure}[h]
    \centering
    \begin{subfigure}[b]{\textwidth}
        \centering
        \includegraphics[width=0.24\linewidth,keepaspectratio=true]{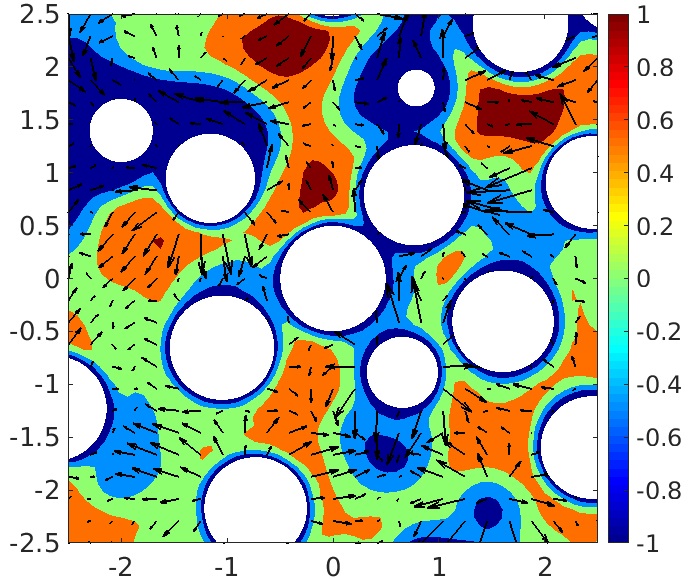}%
        \hfill
        \includegraphics[width=0.24\linewidth,keepaspectratio=true]{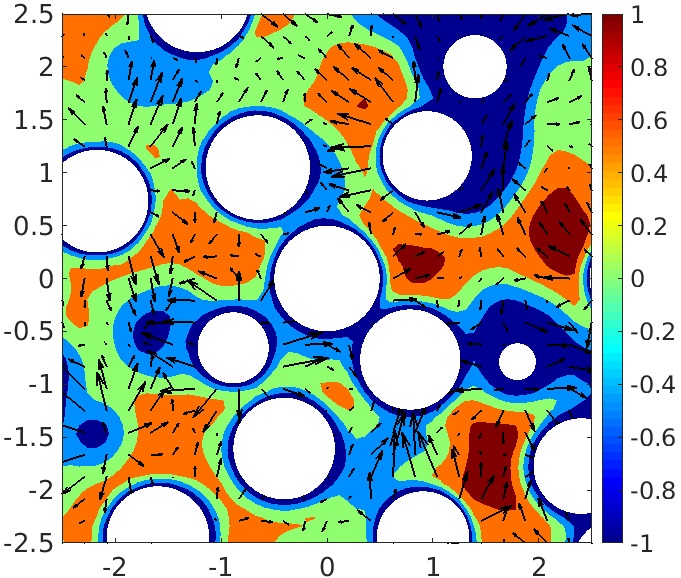}%
        \hfill
        \includegraphics[width=0.24\linewidth,keepaspectratio=true]{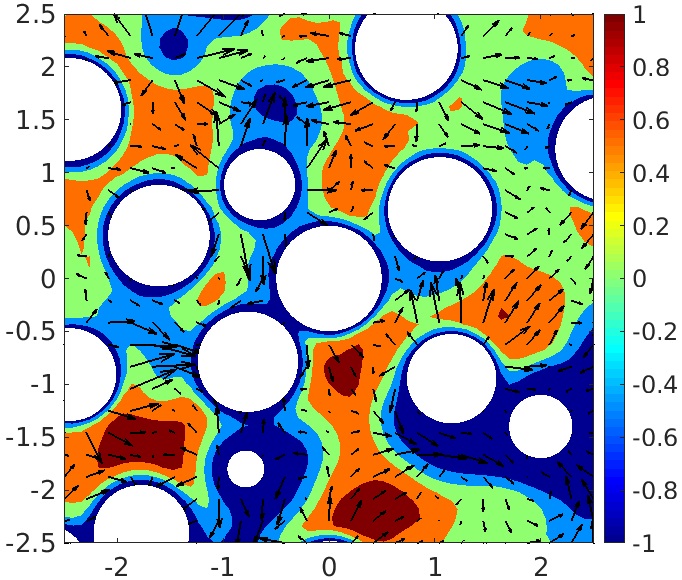}%
        \hfill
        \includegraphics[width=0.24\linewidth,keepaspectratio=true]{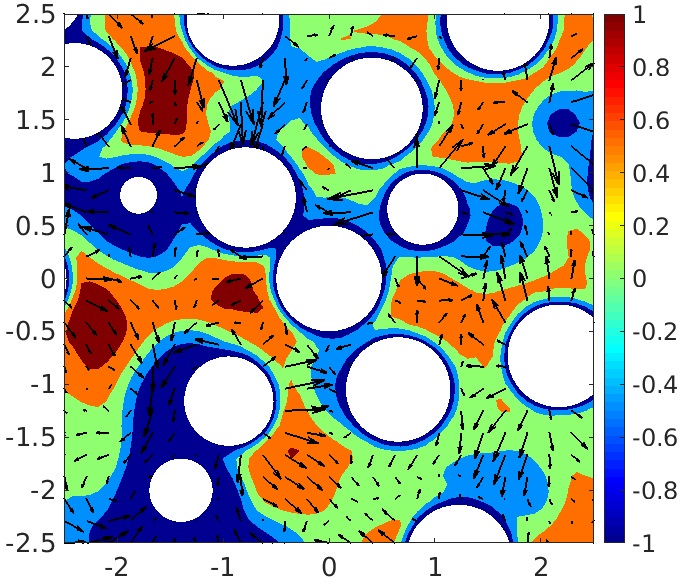}%
        \caption{SE(3)-CNN}
    \end{subfigure}
    \begin{subfigure}[b]{\textwidth}
        \centering
        \includegraphics[width=0.24\linewidth,keepaspectratio=true]{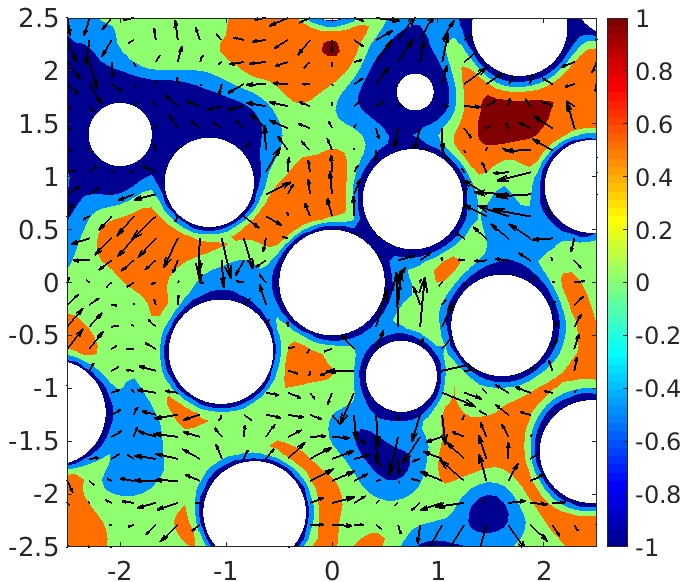}%
        \hfill
        \includegraphics[width=0.24\linewidth,keepaspectratio=true]{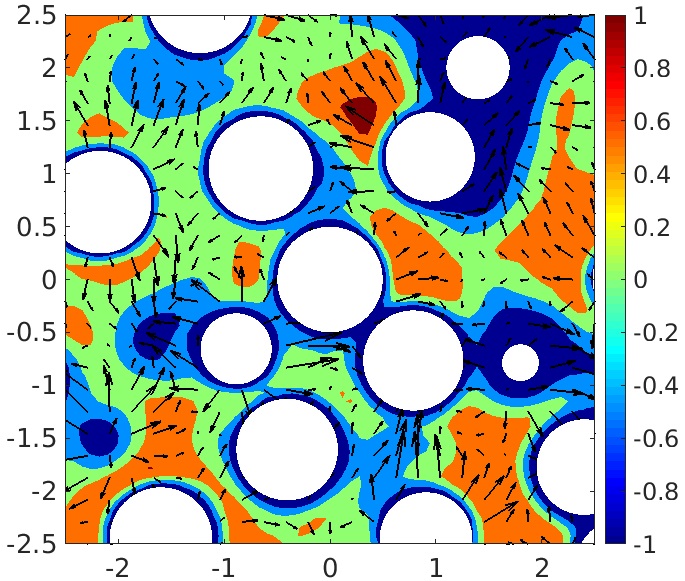}%
        \hfill
        \includegraphics[width=0.24\linewidth,keepaspectratio=true]{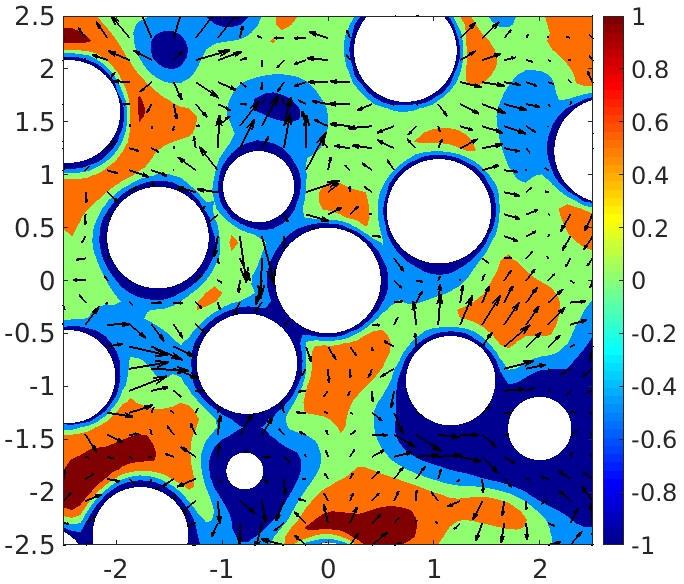}%
        \hfill
        \includegraphics[width=0.24\linewidth,keepaspectratio=true]{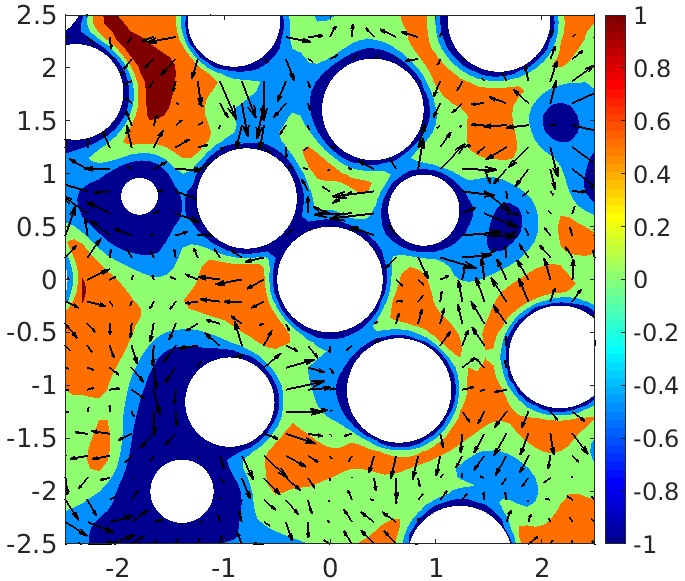}%
        \caption{Data-augmented CNN}
    \end{subfigure}
    \begin{subfigure}[b]{\textwidth}
        \centering
        \includegraphics[width=0.24\textwidth,keepaspectratio=true]{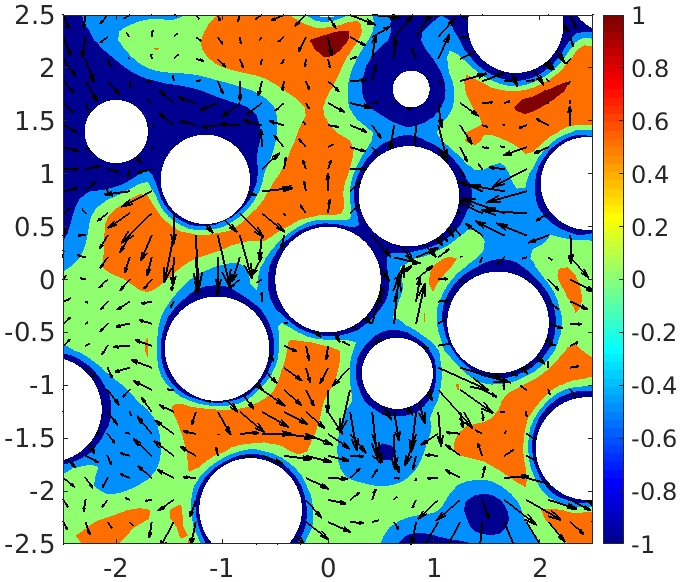}%
        \hfill
        \includegraphics[width=0.251\textwidth,keepaspectratio=true]{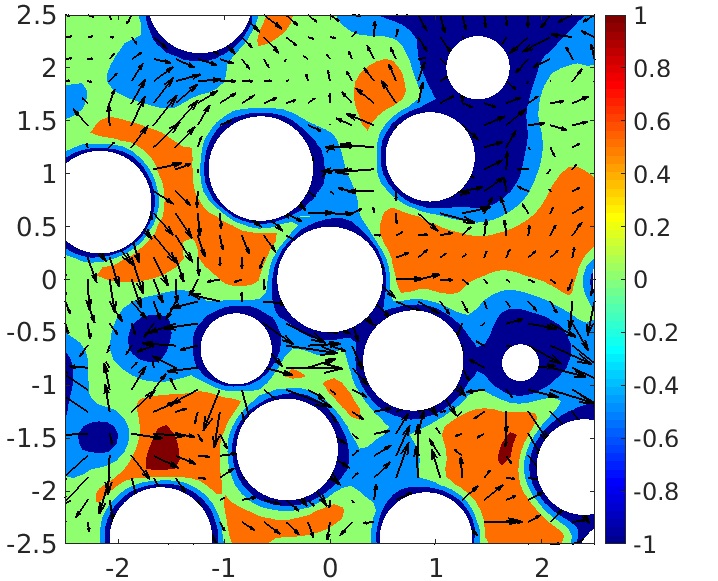}%
        \includegraphics[width=0.24\textwidth,keepaspectratio=true]{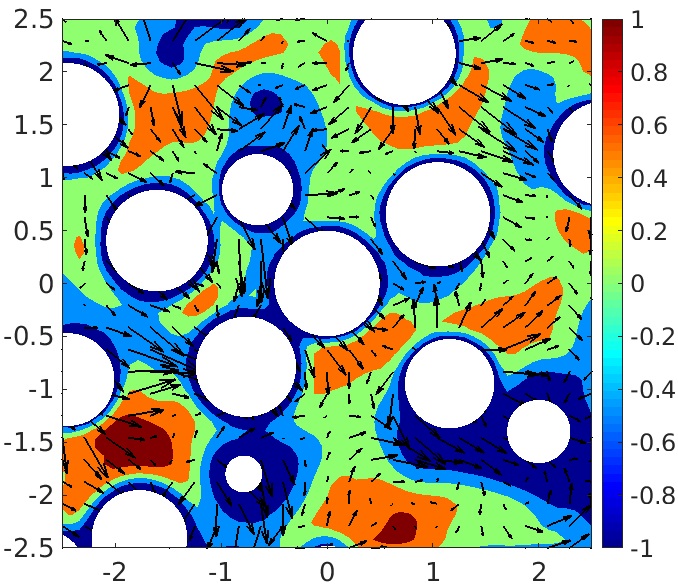}%
        \hfill
        \includegraphics[width=0.24\textwidth,keepaspectratio=true]{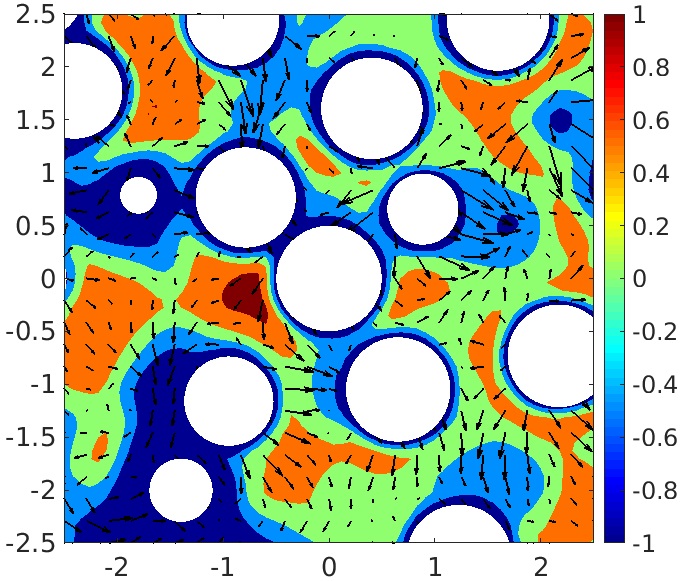}%
        \caption{Simple CNN}
    \end{subfigure}
    \caption{Velocity plots for a central $y-z$ plane of a test sample and its discrete rotations generated using (a) SE(3)-CNN, (b) Data-augmented CNN and (c) Simple CNN from case $Re=86.22$, $\phi=0.21$.}
    \label{fig:v_images}
\end{figure}

Central $y-z$ plane of a random test sample from case-2 has been used here to illustrate the discussed equivariance property of SE(3)-CNN. Pressure and velocity plots on this $y-z$ plane and corresponding $y-z$ planes obtained through discrete $90^0$ degree rotations about the $x$ axis have been shown in Figures \ref{fig:p_images} and \ref{fig:v_images} respectively. The white circular patches of different cross-section correspond to the particles that intersect these planes. For the velocity plots, the streamwise component ($u^{\prime}$) is shown as contour and the other two velocity components ($v^{\prime}, \ w^{\prime}$) are shown as in-plane velocity vector. It can be seen from both the figures, especially the contours of pressure and streamwise velocity component, that the SE(3)-CNN preserves the equivariance property precisely. The Data-augmented CNN yields reasonable results that appear to be equivariant but some differences can be clearly noticed. As expected the least promising equivariant output is produced by Simple CNN.

\subsection{Close Proximity to Reference Particle}
For a given sample, the model is not aware of the presence of neighboring particles that are outside of the $5 \times 5 \times 5$ volume and therefore, the accuracy of the solution will be low close to the outer boundaries of the volume. Furthermore, the accuracy of the flow fields should be higher close to the reference particle as it has deterministic knowledge about all of its immediate neighbors. To confirm the validity of the above reasoning, normalized mean squared error (NMSE) is evaluated for the fluid grid points in $5 \times 5 \times 5$ and $2 \times 2\times 2$ regions around the reference particle. The mathematical definition of these two errors for $u^{\prime}$ are given below as

\begin{equation}
    NMSE_{u^{\prime}, \, (5 \times 5 \times 5)} = \dfrac{\sum\limits_{i=1}^{N_{te}}\sum\limits_{(5\times5\times5)}(u_{DNS}^{\prime}-u_{network}^{\prime})^{2}I_{f}}{\sum\limits_{i=1}^{N_{te}}\sum\limits_{(5\times5\times5)}(u_{DNS}^{\prime})^{2}I_{f}} \, ,
\end{equation}

\begin{equation}
    NMSE_{u^{\prime}, \, (2 \times 2 \times 2)} = \dfrac{\sum\limits_{i=1}^{N_{te}}\sum\limits_{(2\times2\times2)}(u_{DNS}^{\prime}-u_{network}^{\prime})^{2}I_{f}}{\sum\limits_{i=1}^{N_{te}}\sum\limits_{(2\times2\times2)}(u_{DNS}^{\prime})^{2}I_{f}} , .
\end{equation}

These NMSE values are presented in Table \ref{tab:attention_domain}. It can be seen from the results that the error in the $2\times2\times2$ volume is lower than that in the $5\times5\times5$ volume for all three cases. Thus, these results are in agreement with the reasoning that the flow fields are more accurate closer to the reference particle. \citet{siddani2020machine} used this property of accurate flow fields around a reference particle along with voronoi tessellation algorithm to reconstruct flow fields around a given random distribution of stationary particles.

\begin{table}[]
    \centering
    \begin{ruledtabular}
         \begin{tabular}{ccccccc}
         Case & \multicolumn{3}{c}{$(5\times5\times5)$} & \multicolumn{3}{c}{$(2\times2\times2)$} \\
         & $p^{\prime}$ & $u^{\prime}$ & $v^{\prime}, \, w^{\prime}$ & $p^{\prime}$ & $u^{\prime}$ & $v^{\prime}, \, w^{\prime}$ \\
         1 & 0.2467 & 0.3135 & 0.5055 & 0.1444 & 0.2070 & 0.3793 \\
         2 & 0.2197 & 0.1924 & 0.2834 & 0.1178 & 0.1135 & 0.1734 \\
         3 & 0.4059 & 0.1703 & 0.3012 & 0.1751 & 0.0759 & 0.1358
         \end{tabular}
    \end{ruledtabular}
    \caption{Normalized Mean Squared Error (NMSE) for $(5\times5\times5)$ and $(2\times2\times2)$ volumes around reference particle using SE(3)-CNN.}
    \label{tab:attention_domain}
\end{table}

\subsection{Evolution of Loss Function}
\begin{figure}[h]
    \centering
    \begin{subfigure}[b]{0.5\textwidth}
       \includegraphics[width=\textwidth,keepaspectratio=true]{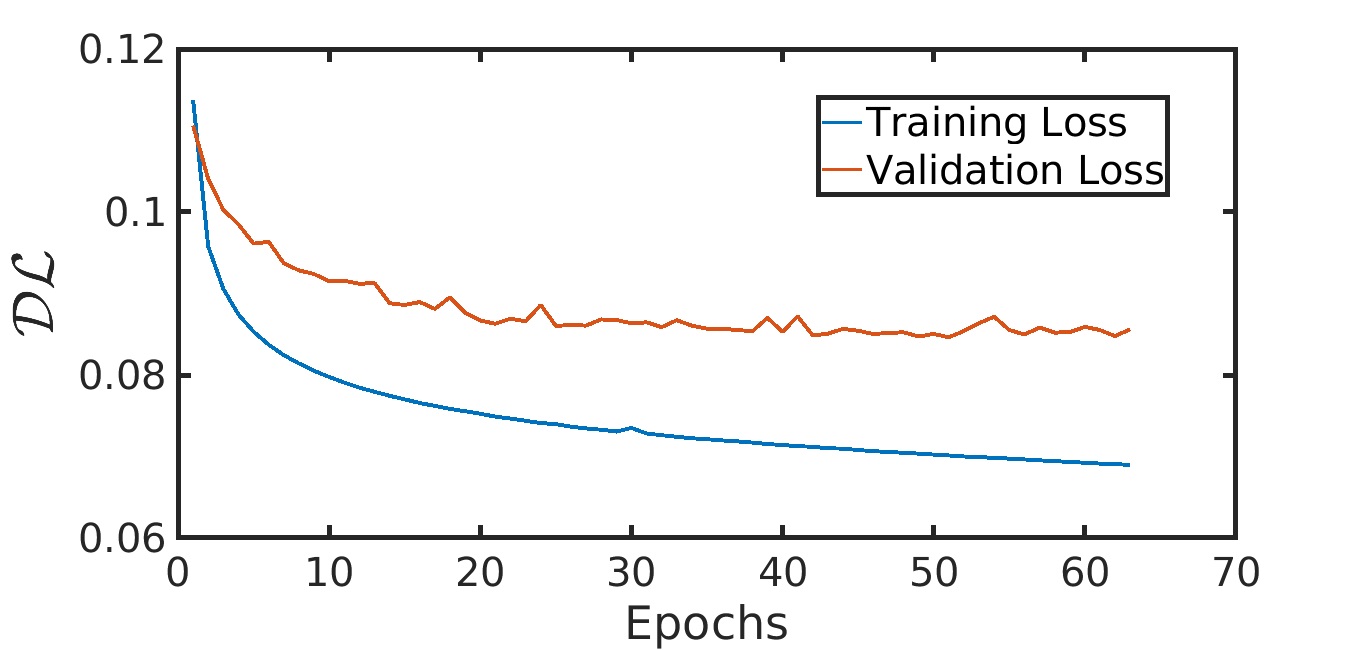}
       \caption{SE(3)-CNN}
    \end{subfigure}%
    \hfill
    \begin{subfigure}[b]{0.5\textwidth}
       \includegraphics[width=\textwidth,keepaspectratio=true]{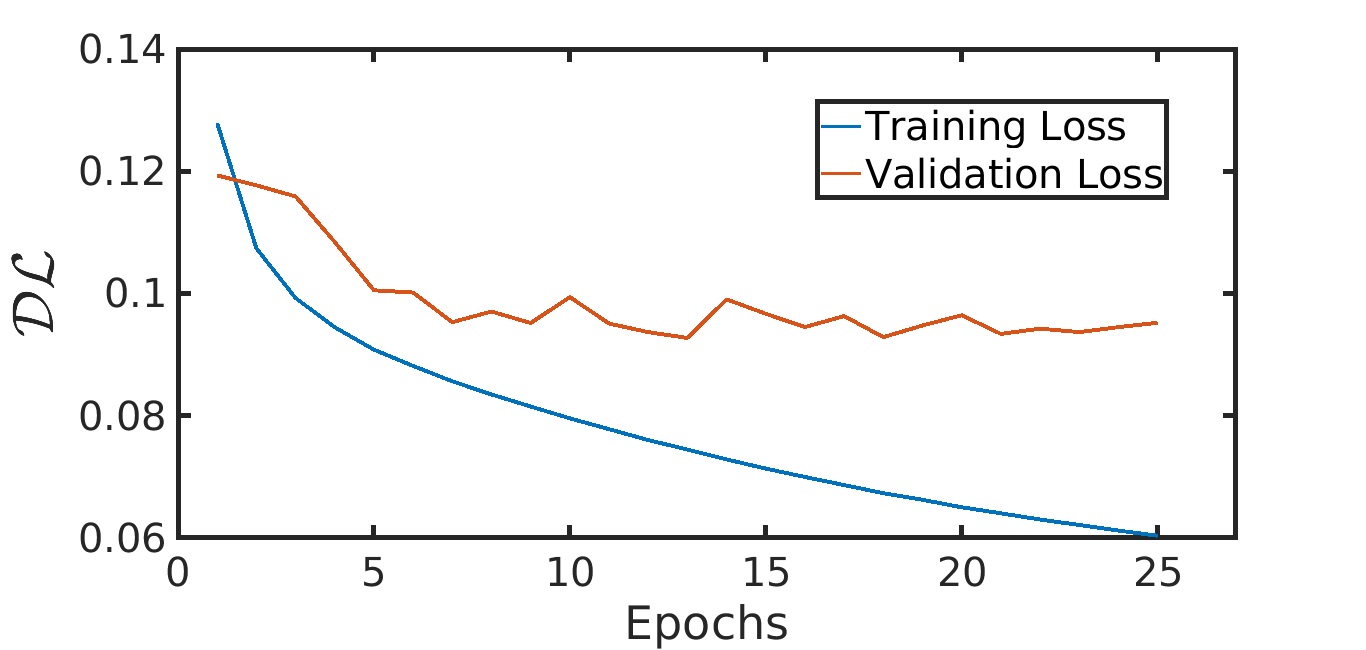}
       \caption{Data-augmented CNN}
    \end{subfigure}%
    \hfill
    \begin{subfigure}[b]{0.5\textwidth}
       \includegraphics[width=\textwidth,keepaspectratio=true]{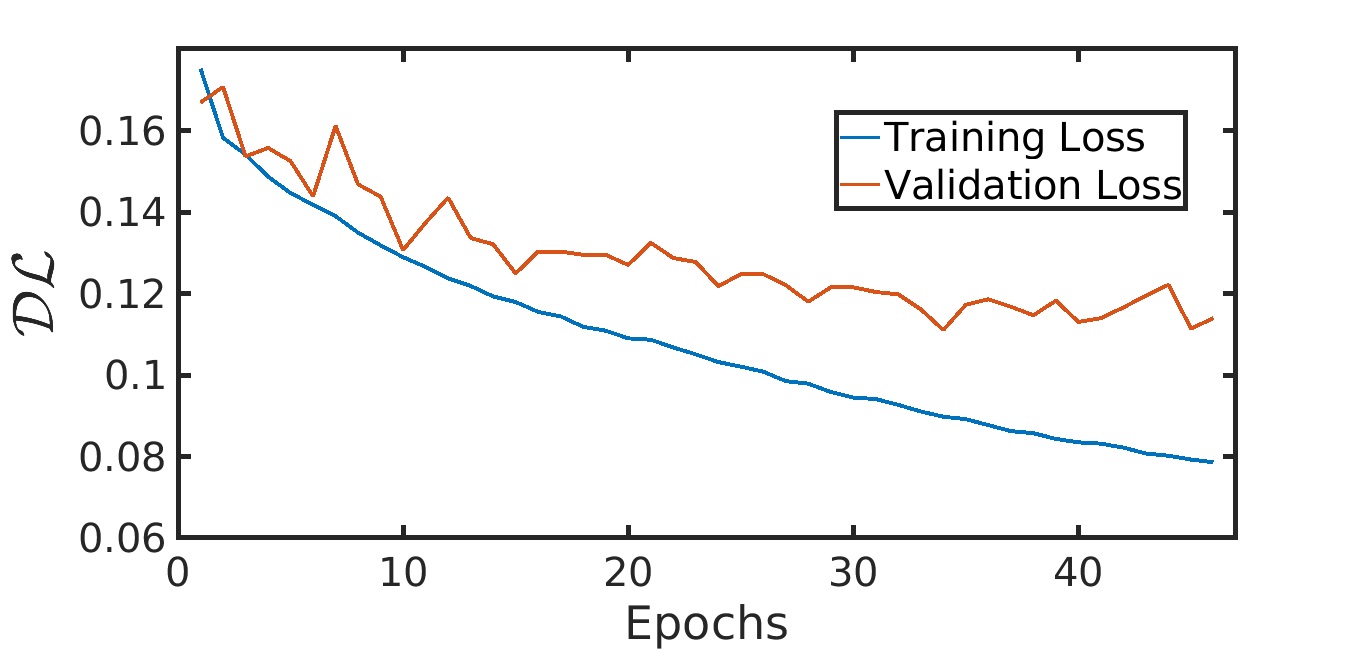}%
       \caption{Simple CNN}
    \end{subfigure}
    \caption{Evolution of Data-based Loss for (a) SE(3)-CNN, (b) Data-augmented CNN and (c) Simple CNN trained on case $Re = 114.60$, $\phi = 0.45$.}
    \label{fig:loss_curve}
\end{figure}

Loss curves for SE(3)-CNN, Data-augmented CNN and Simple CNN trained on case-3 have been presented in Figure \ref{fig:loss_curve}. As data-based loss is the term which is essentially indicative of the network's performance. The decision of \textit{early stopping} of the training process to alleviate the impact of overfitting is taken using the data-based loss. The minimum value of data-based loss for validation dataset, $\mathcal{DL}_{val,min}$, is monitored during the entire training process. The training process is stopped if $\mathcal{DL}_{val,min}$ does not get updated, i.e., data-based loss for validation dataset does not reduce further, in 12 consecutive epochs.

\subsection{Computation Time}
Computational cost with the considered SE(3)-CNN and regular CNN for a mini-batch of size 3 is presented in Table \ref{tab:runtime}. It can be seen from the table that the training time for the SE(3)-CNN is slightly lower than that of the regular CNN. This run-time difference can be associated with the higher number of parameter optimizations that are involved in a regular CNN to update its larger set of independent parameters. However, the testing time for SE(3)-CNN is marginally higher than that of the regular CNN. The higher testing time of SE(3)-CNN can be attributed to the change of basis operation (using $T_{CG}$) involved in the SE(3)-equivariant convolutional layers. This run-time difference can be reduced by pre-storing the kernel/filter in the form of $\hat{\kappa}_{k,j}$ rather than in the form of $\eta_{k,j}$ once the optimized parameters are obtained after the end of training process. This type of pre-storing the kernel eliminates the change of basis operation in SE(3)-convolutional layers during testing process.

\begin{table}[h]
    \centering
    \begin{ruledtabular}
    \begin{tabular}{ccc}
        Computation Type & SE(3)-CNN & Regular CNN \\
        Training & 5.27 s & 5.59 s \\
        Testing & 1.75 s & 1.08 s
    \end{tabular}
    \end{ruledtabular}
    \caption{Computation time in seconds for a mini-batch of size 3 using SE(3)-CNN and regular CNN.}
    \label{tab:runtime}
\end{table}

\subsection{Spatial derivatives of flow fields}
One of the advantages of rotational and reflectional equivariance is their ability to preserve the proper symmetries of tensorial quantities. Although here we have restricted prediction to only pressure (scalar) and velocity (vector) quantities, we can consider tensor quantities constructed out of them, such as gradient of the velocity field. In this section we will consider how well SE(3)-CNN is able to predict these derived tensor quantities. The gradient of flow velocity obtained using SE(3)-CNN, Data-augmented CNN and Simple CNN are produced using second-order accurate central difference scheme at non-boundary fluid grid points. A fluid grid point that satisfies the following criterion is called a non-boundary fluid grid point in this work.
\begin{equation*}
     I_f(\bx + \Delta x \hat{i} \,) + I_f(\bx - \Delta x \hat{i} \,) +
     I_f(\bx + \Delta y \hat{j} \,) +
     I_f(\bx - \Delta y \hat{j} \,) +
     I_f(\bx + \Delta z \,\hat{k} \,) +
     I_f(\bx - \Delta z \,\hat{k} \,)  = 6 \quad.
\end{equation*}
Here, $\Delta x, \ \Delta y, \ \Delta z $ are grid spacing along $x$, $y$, $z$ directions respectively. The derivatives are only evaluated at the non-boundary fluid grid points because second-order accurate central difference scheme can only be applied to these points. Interestingly, the process of obtaining derivatives using a central difference scheme for each variable can be applied as a conventional convolutional layer with the central difference stencil as its kernel.

The normalized mean square error (NMSE) was calculated for each of the nine components of the velcity gradient tensor, and its definition for $\pdv{u^{\prime}}{x}$, is shown below as a example
\begin{equation}
    NMSE_{\pdv{u_{network}^{\prime}}{x}} = \dfrac{\sum\limits_{i=1}^{N_{te}}\sum\limits_{sample}(\pdv{u_{DNS}^{\prime}}{x}-\pdv{u_{network}^{\prime}}{x})^{2}I_{nbf}}{\sum\limits_{i=1}^{N_{te}}\sum\limits_{sample}(\pdv{u_{DNS}^{\prime}}{x})^{2}I_{nbf}} \, ,
\end{equation}
where
\begin{equation*}
    I_{nbf}(x,y,z) = 
    \begin{cases}
        1 & \text{if } (x,y,z)\text{ is a non-boundary fluid point}\\
        0 & \text{else}
    \end{cases}
\end{equation*}
The NMSE was calculated for each of the three networks by comparing their prediction against the corresponding values obtained in PR-DNS, where the gradient was also evaluated using the same central difference scheme. NMSE values related to spatial derivatives of the three networks are presented for case-1 in Table \ref{tab:nmse_derivatives}. Similar to the performance pattern of flow variables, the spatial derivatives of SE(3)-CNN solution have the highest accuracy. The next best accuracy is achieved by Data-augmented CNN.

\begin{table}[]
    \centering
    \begin{ruledtabular}
    \begin{tabular}{*{10}{c}}
       \multicolumn{1}{c}{} & \multicolumn{3}{c}{SE(3)-CNN} & \multicolumn{3}{c}{Data-augmented CNN} & \multicolumn{3}{c}{Simple CNN} \\
         Variable& $\pdv{}{x}$ & $\pdv{}{y}$ & $\pdv{}{z}$ & $\pdv{}{x}$ & $\pdv{}{y}$ & $\pdv{}{z}$ & $\pdv{}{x}$ & $\pdv{}{y}$ & $\pdv{}{z}$ \\
       \hline
       $p^{\prime}$ & 0.2272 & 0.3277& 0.3276& 0.2549& 0.3678& 0.3633& 0.2593& 0.3755& 0.3767\\
       $u^{\prime}$ & 0.3898& 0.3418& 0.3418& 0.4219& 0.3515& 0.3516& 0.4826& 0.4175& 0.4122\\
       $v^{\prime}$ & 0.4098& 0.5186& 0.7575& 0.4573& 0.5529& 0.7816& 0.5657& 0.6599& 0.9058\\
       $w^{\prime}$ & 0.4098& 0.7576& 0.5186& 0.4470& 0.7741& 0.5401& 0.5714& 0.8888& 0.6537
    \end{tabular}
    \end{ruledtabular}
    \caption{Normalized Mean Squared Errors (NMSE) of spatial derivatives for the three networks' predictions are presented for case-1.}
    \label{tab:nmse_derivatives}
\end{table}

It has been mentioned earlier that the SE(3)-CNN ensures the flow information corresponding to transverse directions ($y \, \& \, z$) is statistically the same. This point can be illustrated using the results in Table \ref{tab:nmse_derivatives}. The NMSE values of $\left(\pdv{p^{\prime}}{y} \, \& \pdv{p^{\prime}}{z}\right)$ are very close to each other for SE(3)-CNN solution compared to the Data-augmented CNN or Simple CNN. Even the small difference between these errors using SE(3)-CNN can be attributed to the small deviation of the streamwise direction from $x$-axis in each individual sample, and this leads to the actual transverse directions also slightly deviating from $y \, \& \ z$ axes.

Other pairs that can be compared in the same manner are: $\left(\pdv{u^{\prime}}{y} \, \& \pdv{u^{\prime}}{z}\right)$, $\left(\pdv{v^{\prime}}{y} \, \& \pdv{w^{\prime}}{z}\right)$,
$\left(\pdv{v^{\prime}}{x} \, \& \pdv{w^{\prime}}{x}\right)$ and $\left(\pdv{v^{\prime}}{z} \, \& \pdv{w^{\prime}}{y}\right)$. For all pairs the errors are very small when SE(3)-CNN is used. The errors in case of Data-augmented CNN and Simple CNN are higher indicating their difficulty in replicating statistical property. Preserving the required rotational statistical symmetries of fluid flow around a random distribution of particles is an in-built characteristic property that is preserved by SE(3)-CNN, irrespective of the amount of training data available to it. On the other hand, this statistical nature can only be learned by a regular CNN when sufficient data is available for the training procedure.  The results presented in Table \ref{tab:nmse_derivatives} was when all the training samples (greater than 350) were used in all three approaches. In this case, symmetry preservation of data-augmented and simple CNN were not unacceptable. However, when the same analysis was performed for number of training samples less than 20 or so, the symmetry errors were much larger for the simple CNN, and for the Data-augmented CNN, thus illustrating the importance of SE(3)-CNN in the case of limited data.

\section{Conclusions and Future Scope}
This paper presents a data-driven methodology based on Convolutional Neural Network (CNN) to recreate steady-state particle-resolved flow field around a particle of interest when provided with volume-averaged Reynolds number, mean flow direction and exact locations of neighboring stationary particles in a dispersed multiphase setup. Instead of using conventional CNN, we have implemented a SE(3)-CNN [\onlinecite{weiler20183d}], which is translation and 3D rotation equivariant. This architecture enables incorporation of rotational symmetry. The model only takes a \textit{local} volume that contains the most influential neighbors around a particle of interest to predict the flow field. The results for the considered cases indicate that the equivariant CNN performs better than a conventional CNN with or without data augmentation. This advantage is particularly important in the limit of low availability of training data. While SE(3)-CNN produces very good results even in cases of limited training data, mainly due to implicit enhancement of data due to enforced symmetries, the corresponding results for conventional CNN with or without discrete data augmentation are rather poor. Thus, indicating that incorporating symmetries into data-driven models improves their accuracy, especially under conditions of limited training data.

Since the current work implements rotational symmetry about an arbitrary direction using SE(3)-CNN, this equivariant network can be used for any 3D physical rotation of the problem. SE(3)-CNN's property of intermediate layers being irreducible representations of the SO(3) group is an important step towards interpretable data-driven models. Though the presented methodology has been applied to a steady state problem with stationary particles, it can be easily extended to an unsteady case where mean-flow and particles velocities at current and previous time steps are also taken as inputs to predict flow fields at the current time step.

\begin{acknowledgments}
This work was sponsored by the Office of Naval Research (ONR) as part of the Multidisciplinary University Research Initiatives (MURI) Program, under grant number N00014-16-1-2617. This work was also partly supported and benefited from the U.S. Department of Energy, National Nuclear Security Administration, Advanced Simulation and Computing Program, as a Cooperative Agreement to the University of Florida under the Predictive Science Academic Alliance Program, under Contract No. DE-NA0002378. This work was also partly supported by National Science Foundation under Grant No. 1908299. B.S. would like to thank Mario Geiger and Tess E. Smidt for their helpful suggestions and discussions on the usage of \href{https://github.com/e3nn/e3nn}{e3nn} library [\onlinecite{e3nn_code}].
\end{acknowledgments}

\appendix

\bibliography{references}

\end{document}